\documentclass[amsmath,amssymb,twocolumn,floatfix,aps,showpacs,superscriptaddress,nofootinbib,notitlepage]{revtex4-1}

\usepackage{amsmath,amsthm,amssymb,bm,hyperref,graphicx,color,float,changes,mathtools,enumitem}
\allowdisplaybreaks

\newcommand{\be}{\begin{equation}}
\newcommand{\ee}{\end{equation}}
\newcommand{\bea}{\begin{eqnarray}}
\newcommand{\eea}{\end{eqnarray}}
\newcommand{\6}{\partial}

\newcommand{\inti}{\int_{-\infty}^{+\infty}}

\begin{document}

\title{Quantum critical behavior and thermodynamics of the repulsive one-dimensional Hubbard model in a magnetic field}

\author{Ovidiu I. P\^{a}\c{t}u}
\affiliation{Institute for Space Sciences, Bucharest-M\u{a}gurele, R 077125, Romania}
\author{Andreas Kl\"umper}
\affiliation{Fakult\"at f\"ur Mathematik und Naturwissenschaften, Bergische
 Universit\"at Wuppertal, 42097 Wuppertal, Germany}
\author{Angela Foerster}
\affiliation{Instituto  de  F\'{i}sica  da  UFRGS,  Av.  Bento Gon\c{c}alves  9500,  Porto  Alegre,  RS,  Brazil}

\begin{abstract}

Even though the Hubbard model is one of the most fundamental models of highly
correlated electrons, analytical and numerical data describing its
thermodynamics at nonzero magnetization are relatively
scarce. We present a detailed investigation of the thermodynamic properties
for the one dimensional repulsive Hubbard model in the presence of an
arbitrary magnetic field for all values of the filling fraction and
temperatures as low as $T \sim 0.005\, t.$ Our analysis is based on the system
of integral equations derived in the quantum transfer matrix framework. We
determine the critical exponents of the quantum phase transitions and also
provide analytical derivations for some of the universal functions
characterizing the thermodynamics in the vicinities of the quantum critical
points. Extensive numerical data for the specific heat, susceptibility,
compressibility, and entropy are reported. The experimentally relevant double
occupancy presents an interesting doubly nonmonotonic temperature dependence
at intermediate values of the interaction strength and also at large repulsion
and magnetic fields close to the critical value. The susceptibility in zero
magnetic field has a logarithmic singularity at low temperatures for all
filling factors similar to the behavior of the same quantity in the spin-1/2 isotropic
Heisenberg model. We determine the density profiles for a harmonically trapped system and
show that while the total density profile seems to depend mainly on the value
of chemical potential at the center of the trap the distribution of phases in
the inhomogeneous system changes dramatically as we increase the magnetic
field.

\end{abstract}

\maketitle

\section{Introduction}

Originally introduced to describe interacting electrons in narrow energy bands
the Hubbard model \cite{Hubbard1, Hubbard2,Gutzw,Kana,EFGKK} is a paradigmatic
model of condensed matter physics which has been used to describe the Mott
metal-insulator transition, band magnetism and is believed to capture
essential physics of high-$T_c$ superconductivity.

Ultracold fermions in optical lattices provide realizations of the Hubbard
model in a very clean environment with a high degree of control over
temperature, chemical potential, on-site repulsion, tunneling amplitude and
even dimensionality. Very recently, experimental realizations of the repulsive
one-dimensional (1D) Hubbard model were reported together with observations of
antiferromagnetic spin correlations \cite{Gross1,Gross2}, incommensurate
magnetism \cite{Gross3} and measurement of non-equilibrium transport
properties \cite{SKHSB}.  The many-body physics of 1D systems presents
features which distinguish them from their higher-dimensional counterparts. In
one dimension Fermi liquid theory breaks down and for many systems the
appropriate low energy description is given by the Tomonaga-Luttinger liquid
(TLL) theory \cite{Hald} which predicts counterintuitive phenomena like
spin-charge separation in multi-component systems \cite{Giamar}. In addition,
many relevant models are integrable and the exact solutions
of such models play an important role in the study of non-perturbative effects
in strongly correlated systems.  Fortunately, the 1D Hubbard model belongs to
this class of systems and can be solved using the nested Bethe ansatz
\cite{LW}. At zero temperature a relatively large body of knowledge has been
accumulating steadily including: elementary excitations
\cite{Ovch,UF,Coll,Woyn1,Woyn2,Woyn3,KSZ,EK,DEGKEK}, complete set of
eigenstates \cite{EKS}, magnetic properties \cite{Tak2,Tak3,Shiba3,CHO},
symmetries \cite{Yang2,UK,MG,MG1} and correlation functions \cite{OS,
  OSS,Schulz2,Schulz1,FK1,FK2,PS,PS1,SMil,WSTS,Weng,PMS,PHMS1,PHMS2,GM,GBSTK,IPA1,AIP,CZ,BTC,CBP,CBP1,CPSSC,
  BCPS,CC1,CC2,Essl1,SEPSV,CNP}.

In this paper we study the properties of the Hubbard model at finite
temperature. The first thermodynamic description of the model was derived
shortly after the Lieb and Wu solution by Takahashi \cite{Tak6} assuming the
string hypothesis and using the thermodynamic Bethe ansatz (TBA)
\cite{Tak1}. Unfortunately, in this case TBA produces an infinite number of
non linear integral equations, which are extremely important in the study of
low temperature properties, but very hard to implement numerically. It took
almost twenty years until a numerical implementation, with reasonable
accuracy, of this system of equations was reported in the literature
\cite{KUO1, UKO2,UKO1,TS}. Certain simplifications appear in the strong
coupling limit and in the spin-disordered regime \cite{Klein,Ha,EEG,VMC}.  A
different system of equations describing the thermodynamics of the Hubbard
model for all filling fractions was derived in \cite{JKS} making use of the
quantum transfer matrix (QTM) \cite{Suz,SI,Koma,SAW,K1,K2} (at half filling
similar equations were derived in \cite{Tsuneg,KBar}). This description
involves only six auxiliary functions and while the numerical implementation
is also nontrivial accurate numerical data can be obtained for almost all
values of the relevant physical parameters. Another advantage of the QTM is
that it also allows for the investigation of some correlation functions at
finite temperature \cite{Tsuneg,KBar,USK}.  The complexity of both
thermodynamic descriptions resulted in relatively few results in the
literature for the temperature dependence of various thermodynamic quantities,
especially in the presence of a magnetic field. Recent experimental
realizations of the repulsive 1D Hubbard model
\cite{Gross1,Gross2,Gross3,SKHSB} involve both spin balanced and imbalanced
systems and temperatures in the range of $T\in [0.2\,t, 1.5\,t]$ for which
thermal effects are extremely significant. For these reasons in this paper we
perform a detailed study of the thermodynamic properties and quantum critical
behavior of the model in a magnetic field and for a wide interval of
temperatures: $T\in [0.005\,t,3\,t]$.

At zero temperature the phase diagram of the Hubbard model is
very rich with a multitude of quantum phase transitions
(QPTs) induced by either the variation of the chemical potential or magnetic
field. In the vicinity of the quantum critical points the thermodynamics is
expected to be universal and completely characterized by the universality
class of the transition \cite{Sachdev}. Our analysis of 6 QPTs showed that in
all cases the critical exponents are $z=2$ and $\nu=1/2$ but the universal
thermodynamics is not necessarily the one of free fermions.  For example, the
transitions from the vacuum to the non half filled system with zero
magnetization (at zero magnetic field) or fully polarized system (at non zero
magnetic field) belong to the universality class of the spin-degenerate
impenetrable particle gas \cite{PKF} for which the universal thermodynamics is
described by Takahashi's formula (\ref{tt1}).  The dependence of the specific
heat, magnetic susceptibility, compressibility, entropy, and double occupancy
on the magnetic field is extremely complex at low temperatures. Particularly
interesting is the presence of two minima in the dependence on temperature of
the double occupancy as a result of the competition between charge and spin
modes and related to the Pomeranchuk effect \cite{STUBG}.  The magnetic
susceptibility at zero magnetic field presents a logarithmic dependence on the
temperature for all filling fractions similar to the case of the spin-1/2 isotropic Heisenberg model (XXX spin
chain) \cite{EAT,K3}. We show that the interval of temperatures for which the
TLL description is valid (linear dependence of the specific heat) decreases at
lower filling fractions and also as the coupling strength
increases. Experimental realizations in optical lattices require the presence
of a confining parabolic potential which breaks the integrability of the
model. In this case the solution of the homogeneous model
coupled with the local density approximation can provide accurate results in
the case of slowly varying potentials and large number of particles. The
density profiles computed in this way show that while the total density
profile is almost unchanged as we increase the magnetic field the distribution
of phases present in the inhomogeneous system is extremely sensitive.

The plan of the paper is as follows. In Sect.~\ref{s1} we introduce the
Hubbard model and review the properties of the ground state in
Sect.~\ref{s2}. The thermodynamic description obtained in the QTM framework is
presented in Sect.~\ref{s3}.  In Sect.~\ref{s4} we investigate the quantum
critical behavior and the universal thermodynamics in the vicinity of the
quantum critical points. Detailed results for the specific heat, magnetic
susceptibility and compressibility at below half filling are reported in
Sect.~\ref{s5} and in the half filled case in Sect.~\ref{s6}. Sections
\ref{s7} and \ref{s8} contain the dependence of the double occupancy and
entropy on filling fraction and magnetic field. An analysis of the density
profiles of the model in the presence of a trapping potential using the local
density approximation is presented in Sect.~\ref{s9}. We conclude in
Sect.~\ref{s10}. Some technical details regarding the numerical implementation
of the two types of convolutions appearing in the QTM
equations can be found in Appendices~\ref{app1} and \ref{app2}.

\section{The Hubbard model}\label{s1}

The one-dimensional Hubbard model describes interacting fermions on a lattice. Assuming an arbitrary  magnetic
field the Hamiltonian is
\begin{align}\label{ham}
\mathcal{H}=\mathcal{H}_{kin}+\mathcal{H}_{int}+\mathcal{H}_{ext}\, ,
\end{align}
where
\begin{subequations}\label{hamc}
\begin{align}
\mathcal{H}_{kin}&=-t \sum_{j=1}^L\sum_{a=\{\uparrow,\downarrow\}}
\left(c_{j+1,a}^\dagger c_{j,a}+c_{j,a}^\dagger c_{j+1,a}\right)\, ,\\
\mathcal{H}_{int}&=U\sum_{j=1}^L \left(n_{j,\uparrow}-\scalebox{0.8}{$\frac{1}{2}$}\right)\left(n_{j,\downarrow}-\scalebox{0.8}{$\frac{1}{2}$}\right)\, ,   \\
\mathcal{H}_{ext}&=-\sum_{j=1}^L \left[ \mu(n_{j,\uparrow}+n_{j,\downarrow})+H (n_{j,\uparrow}-n_{j,\downarrow})\right]\, .
\end{align}
\end{subequations}
In the defining relations (\ref{hamc}) $L$ is the number of lattice sites of
the system, $c_{j,a}^\dagger$ and $c_{j,a}$ are creation and annihilation
operators of an electron of spin $a$ ($a\in\{\uparrow,\downarrow\}$) at site
$j$ of the lattice and $n_{j,a}=c_{j,a}^\dagger c_{j,a}$. The operators
$c_{j,a}^\dagger$ and $c_{j,a}$ are Fermi operators and satisfy canonical
anticommutation relations $
\{c_{j,a},c_{k,b}\}=\{c_{j,a}^\dagger,c_{k,b}^\dagger\}=0\, ,$
$\{c_{j,a},c_{k,b}^\dagger\}=\delta_{j,k}\delta_{a,b}\, $ with
$j,k\in\{1,\cdots,L\}\, $ and $a,b \in\{\uparrow,\downarrow\}.$
The two real numbers $t$ and $U$ quantify the strength of the
tight-binding and the Coulomb interaction terms. Due to the
various symmetries of the Hubbard model we may map systems with $U<0$ to $U>0$
and $\mu\ge 0$ to $\mu\le 0$, see (\ref{symmHmu},\ref{symm}). Hence, without
loss of generality we focus on the case of repulsive interaction between the
electrons ($U > 0$) and electron densities between 0 and 1 (``half filling'').
Also, we measure energies in units of $t$ which is equivalent to setting $t =
1$.  We will also set $k_B=\mu_B=1$ in the rest of the paper. We restore the units in the
figures and their captions.  Finally, $\mu$ and
$H$ are the chemical potential and magnetic field.

\section{Thermodynamics at zero temperature}\label{s2}

The main goal of this paper is the investigation of the quantum critical
behavior and thermodynamic properties of the repulsive Hubbard model in a
magnetic field.  First, it is useful to remind of the
description of the ground state and the phase diagram at zero temperature
which allows for the identification of the quantum phase
transitions. We will also review the charge and spin velocities and the known
analytical formulae for the ground state susceptibilities.

The ground state of the model is described by a system of integral equations for root densities first obtained
in \cite{LW}:
\begin{subequations}\label{densities}
\begin{align}
\rho(k)&=\frac{1}{2\pi}+\cos k\int_{-A}^A d\lambda\,  a_1(\sin k-\lambda)\sigma (\lambda)\, ,\\
\sigma(\lambda)&=\int_{-Q}^Q dk\, a_1(\lambda-\sin k) \rho(k)\nonumber\\
&\ \ \ \ \ \ \ \ \ \ \ \ \ \ \ \ \ \ \
-\int_{-A}^A d\lambda'\, a_2(\lambda-\lambda')\sigma(\lambda')\, ,
\end{align}
\end{subequations}
with the kernels defined by
\be
a_l(x)=\frac{1}{2\pi}\frac{2 l u}{(l u)^2+x^2}\, , \ \ \  u=\frac{U}{4 t}\, .
\ee
For a system of $N$ electrons of which $M$ have spin down  the parameters $Q$ and $A$ fix
the particle density $n$  and magnetization per site $m$ via
\begin{align}
n=&\frac{N}{L}=\int_{-Q}^Q dk\, \rho(k)\, ,\label{d}\\
m=&\frac{N-2M}{2L}=\frac{1}{2}\left[\int_{-Q}^Q dk\, \rho(k)-2
\int_{-A}^A d\lambda\, \sigma(\lambda)\right]\, ,\label{m}
\end{align}
and the ground state free energy per site is ($e$ is the energy per site)
\begin{align}
f&= e-\mu n-2 H m\, , \nonumber \\
 &=\int_{-Q}^Qdk\, (-2\cos k -\mu-2u-H)\rho(k)\nonumber\\
&\ \ \ \ \ \ \ \ \ \ \ \ \ \ \ \  +2H\int_{-A}^Ad\lambda\, \sigma(\lambda)+u\, .
\end{align}
The dressed energies satisfy the following system of integral equations
\begin{subequations}\label{dressede}
\begin{align}
\bar\kappa(k)&=-2\cos k -\mu-2 u -H\nonumber\\
&\ \ \ \ \ \ \ \ \ \ \ \ \ \ \ \ \ \ +\int_{-A}^A d\lambda\,a_1(\sin k -\lambda)\varepsilon(\lambda)\, ,\\
\varepsilon(\lambda)&=2H+\int_{-Q}^Q dk\,  \cos k\, a_1(\sin k-\lambda)\bar\kappa(k)\nonumber\\
&\ \ \ \ \ \ \ \ \ \ \ \ \ \ \ \ \ \ -\int_{-A}^A d\lambda' a_2(\lambda-\lambda')\varepsilon(\lambda')\, ,
\end{align}
\end{subequations}
and they play an important role in the investigation of the phase diagram at zero temperature. In addition
to  fixing the particle density and magnetization  the integration boundaries are also the points at which
the dressed energies switch sign. As functions of the chemical potential and magnetic field they are determined
from the conditions
\be
\bar\kappa(\pm Q)=0\, ,\ \ \ \ \ \varepsilon(\pm A)=0\, .
\ee

\subsection{Ground state phase diagram}\label{s21}

\begin{figure}
\includegraphics[width=\linewidth]{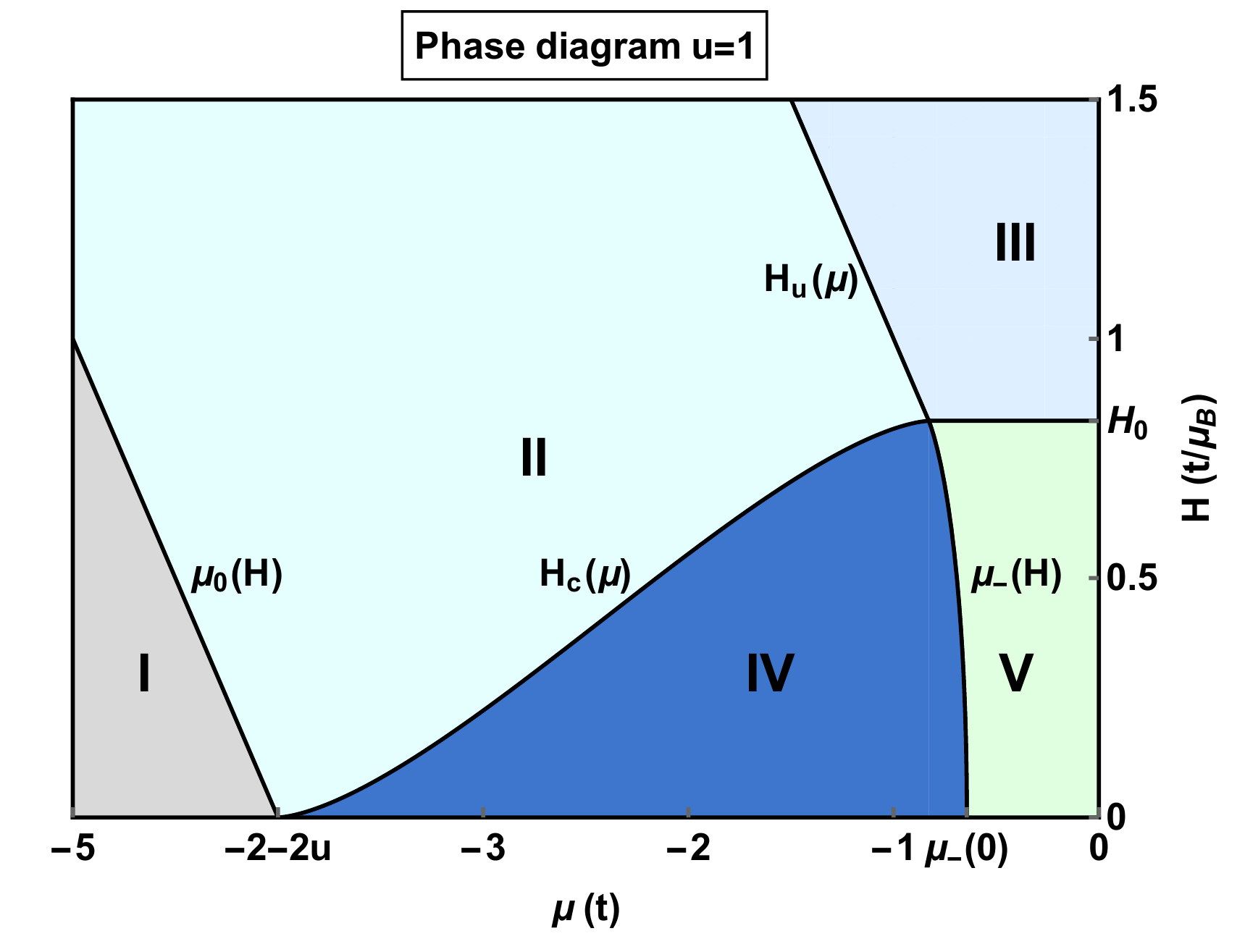}
\caption{Ground state phase diagram in $\mu-H$ coordinates for $u=1$. The value of
$\mu_-(0)$ is given by Eq.~(\ref{muminus}) and $H_0(u)= 2\left(1+u^2\right)^{1/2}-2u$.
In terms of the densities and magnetization the five phases are characterized by  I: $ n=m=0$ ,
 II: $ n_\uparrow>0\, ,n_\downarrow=0$,  III: $ n_\uparrow=1\, , n_\downarrow=0$,  IV: $0<n<1\, , m\ge0$
and  V: $ n=1\, ,m\ge0$.
 }
\end{figure}

At zero temperature and $\mu\le 0\, ,H\ge 0$ the Hubbard model presents five
phases \cite{Tak7,EFGKK} and we should point out that due to the particular
form of the Hamiltonian (\ref{ham}) the system is at half filling ($n=1$) for
$\mu=0$. For a fixed value of the magnetic field by varying the chemical
potential the Hubbard model presents quantum phase transitions at the quantum
critical points (QCPs). The universality classes of these QPTs and the
associated universal thermodynamics in the vicinity of the QCPs will be
studied in Sect.~\ref{s4}.  The phases can be distinguished by
  the densities of spin-up and spin-down electrons $n_\uparrow, n_\downarrow$
  which serve as order parameters.  The five phases and boundaries (which are
critical lines of QCPs) are:

\begin{itemize}[leftmargin=*]

\item {\bf Phase I} $(n=m=0)$: Vacuum.
This phase is characterized by zero density of electrons and $Q=A=0.$ The boundary is
defined by the condition
\be
\mu\le\mu_0(H)=-2-2u-H\, .
\ee
Clearly, all correlation functions are trivial.
\item {\bf Phase II} $(n_\uparrow>0\, ,n_\downarrow=0)$: Partially filled, spin polarized band. The total
density is $0<n<1$ and the magnetization $m=n/2$. In this phase the parameter $Q$  can take values between
$0$ and $\pi$ and $A=0$. The chemical potential
is related to $Q$ and $H$ via
\[
\cos Q=-\frac{1}{2}\left(\mu+H+2u\right)\, ,
\]
and the  magnetic field satisfies $H_c\le H\le H_u$
with
\begin{subequations}
\begin{align}
H_c&=\frac{2 u}{\pi}\int_{0}^Qdk\, \cos k \frac{\cos k-\cos Q}{u^2+(\sin k)^2}\, ,\\
H_u&=2-\mu-2u\, .
\end{align}
\end{subequations}
The density is given by $n=\arccos\left(1-\frac{\mu-\mu_0(H)}{2}\right)/\pi$.

Here, the density-density correlation function and the
  one-particle Green's function decay algebraically with free fermion
  exponents, the spin-spin correlation functions are trivial.
\item  {\bf Phase III} $(n_\uparrow=1\, , n_\downarrow=0)$: Half filled, spin polarized band. The total density is
$n=1$ and the magnetization is $m=1/2$. The boundaries
are given by  $ (Q=\pi\, , A=0)$

\begin{subequations}
\begin{align}
H \ge&\,   H_0(u)=2\left(1+u^2\right)^{1/2}-2 u\, ,\label{H0}\\
\mu\ge&\, 2-2u-H\, .
\end{align}
\end{subequations}
The correlation functions are frozen. Note that the boundaries of phase III in the $ \mu-H$ plane are straight lines,
the one described by (\ref{H0}) is strictly horizontal.
\item  {\bf Phase IV} $(0<n<1\, , m\ge0)$: Partially filled and magnetized band.
  In this phase $0<Q<\pi\, ,0<A\le \infty.$

The density-density and spin-spin correlation functions
  as well as the Green's functions decay algebraically.

\item  {\bf Phase V} $(n=1\, ,m\ge0)$: Half filled and magnetized band. In this
region $Q=\pi$  and $0<A\le \infty$. At $H=0$ the value of the chemical potential $\mu_-$
which separates phase IV and phase V is given by
\be\label{muminus}
\mu_-(H=0)=2-2u-2 \int_0^\infty \frac{d\omega}{\omega}\frac{ J_1(\omega) e^{-\omega u}}{\cosh (\omega u)}\, .
\ee
For $H>0$ the boundary between phase IV and phase V is determined by the condition $\bar{\kappa}(\pm \pi)=0$
(in Eqs.~(\ref{dressede}) we set $Q=\pi$).
As only charge neutral excitations are gapless,  the
  density-density and spin-spin correlation functions decay algebraically,
  the one-particle Green's function decays exponentially.
\end{itemize}
The exponents of the algebraic decay of correlation functions
  in phases IV and V are given by conformal weights resp. Luttinger liquid
  parameters that are obtained \cite{EFGKK} from the dressed charge discussed
  below in subsection \ref{dressedcharge}.

\begin{figure}[t]
\includegraphics[width=\linewidth]{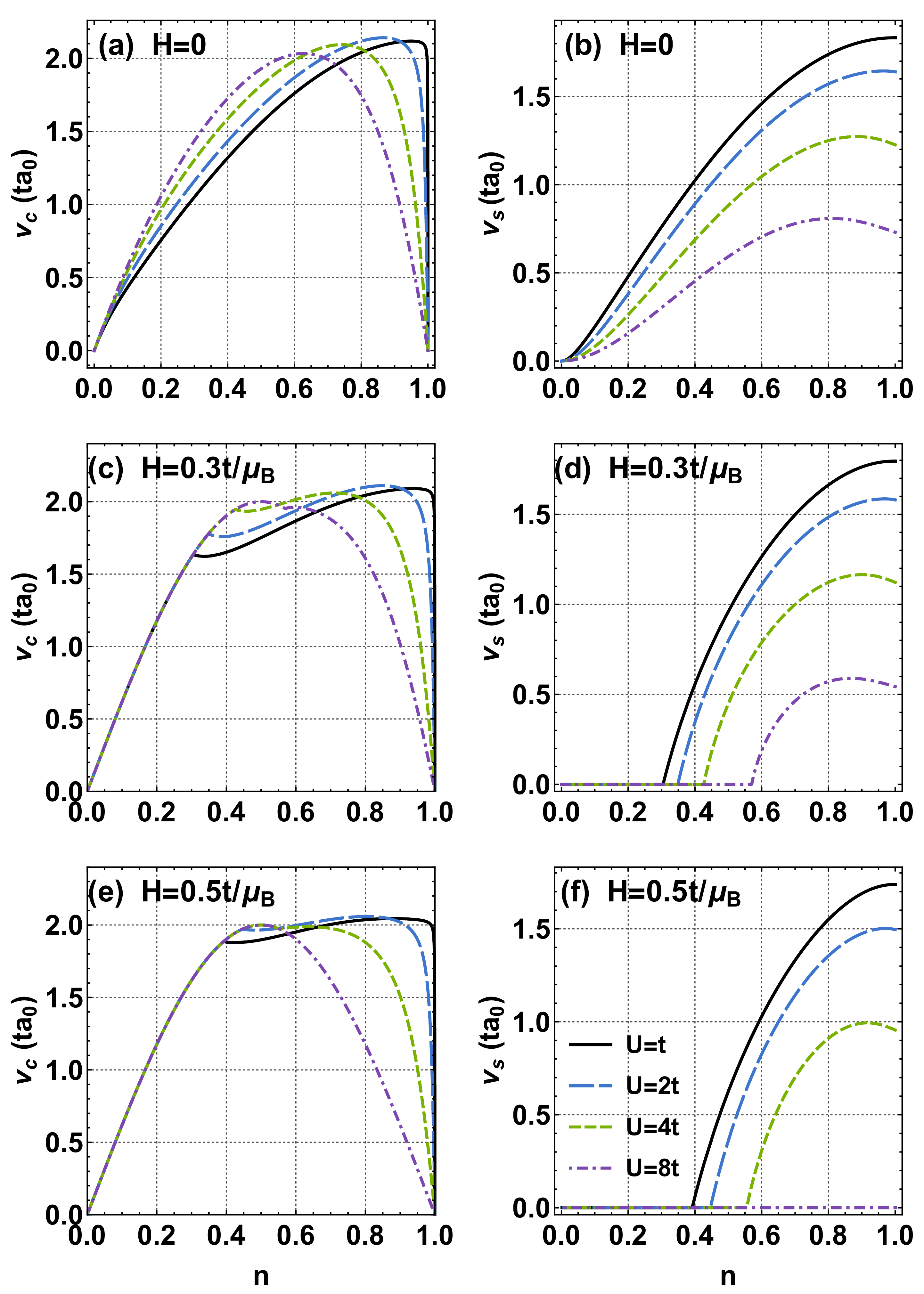}
\caption{Density dependence of the charge and spin velocities ($a_0$ is the lattice spacing) for $H=0$ (upper panels), $H=0.3\, t/\mu_B$
(middle panels) and $H=0.5\,t/\mu_B$ (bottom panels). }
\label{velocities}
\end{figure}

\subsection{Spin and charge velocities}

The charge and spin velocities can be calculated from
\be\label{vcs}
v_c=\left.\frac{\bar\kappa'(k)}{2\pi\rho(k)}\right|_{k=Q}\, , \ \ \
v_s=\left.\frac{\varepsilon'(k)}{2\pi\sigma(\lambda)}\right|_{\lambda=A}\, ,
\ee
where the derivatives of the dressed energies satisfy the following system of integral
equations
\begin{align*}
\bar\kappa'(k)&=2\sin k +\cos k \int_{-A}^A d\lambda\,a_1(\lambda-\sin k)\varepsilon'(\lambda)\, ,\\
\varepsilon'(\lambda)&=\int_{-Q}^Q dk\,  \, a_1(\lambda-\sin k)\bar\kappa'(k)\nonumber\\
&\ \ \ \ \ \ \ \ \ \ \ \ \ \ \ \ \ \ \ \ \ \  \   -\int_{-A}^A d\lambda' a_2(\lambda-\lambda')\varepsilon'(\lambda')\, .
\end{align*}
\begin{figure}
\includegraphics[width=\linewidth]{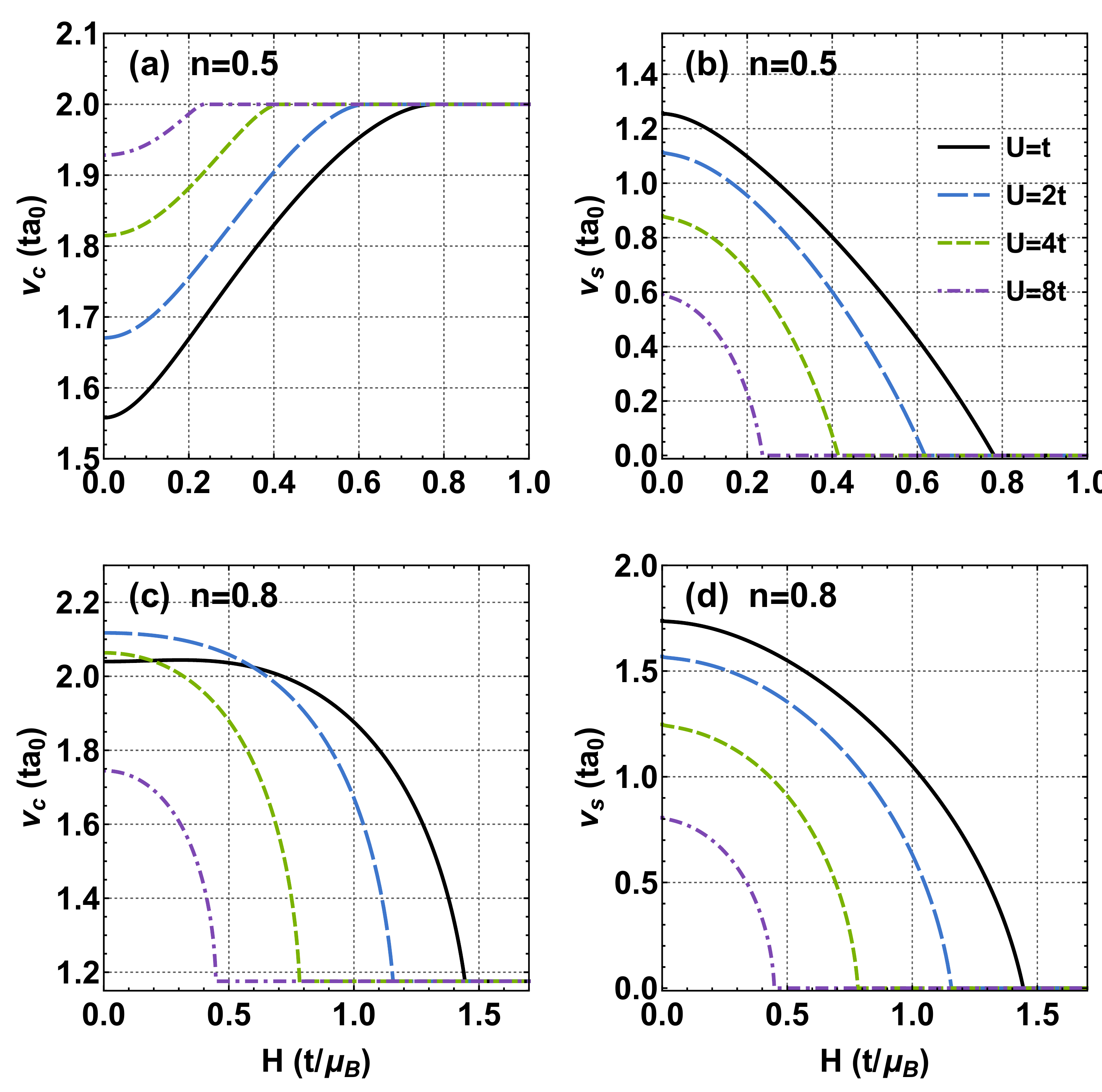}
\caption{Magnetic field dependence of the charge and spin velocities for two fixed values of the
density $n=0.5$ (upper panels) and $n=0.8$ (lower panels). For large values of the magnetic field
the charge velocities are $v_c(n)=2 \sin (\pi n).$}
\label{velocitiesh}
\end{figure}
The charge and spin velocities play an important role in the low temperature description of the
Hubbard model. In this regime certain phases are described by the TLL theory and
the free energy takes the form \cite{Hald,Giamar,Affleck}
\begin{subequations}
\begin{align}
f_{II}&=e_{II}-\frac{\pi T^2}{6}\frac{1}{v_c}\, ,\ \ &\mbox{phase II}\, ,\\
f_{IV}&=e_{IV}-\frac{\pi T^2}{6}\left(\frac{1}{v_c}+\frac{1}{v_s}\right)\, ,\ \ &\mbox{phase IV}\, ,\\
f_{V}&=e_{V}-\frac{\pi T^2}{6}\frac{1}{v_s}\, ,\ \ &\mbox{phase V}\, .
\end{align}\label{lowT}
\end{subequations}
Therefore, the entropy and the specific heat in  phases II, IV and V have a linear dependence
on temperature for temperatures close to zero and the slope (also known as the specific heat coefficient)
can be determined from the knowledge of the velocities.

The dependence  of the velocities on the filling factor at zero and nonzero magnetic field is presented
in Fig.~\ref{velocities}. To our knowledge the only result reported in the literature (see \cite{Schulz1}
and Chap.~VI of \cite{EFGKK}) is the case of  zero magnetic field which is presented in the panels
(a) and (b) of Fig.~\ref{velocities}. The charge  velocity is zero at both $n=0$ and $n=1$ signalling
the quantum phase transitions between phases I and IV  and IV and V (for $H=0$ and $0<n<1$ the system
is in phase IV). For weak interactions $v_c$ presents  a rapid variation close to half filling. The spin
velocity is zero only at $n=0$ and reaches a finite  value at $n=1$.

The presence of a magnetic field introduces additional complications. For small values of $n$ and $H>0$
the system is fully polarized (phase II, $A=0$) and from  Eqs.~(\ref{densities}) and (\ref{dressede})
we have $\rho(k)=1/2\pi$ and $\bar\kappa(k)=-2\cos k-\mu-2 u -H$. Together with Eq.~(\ref{d}) and the
definitions (\ref{vcs}) they imply that in phase II $v_c=2 \sin (\pi n)$ and $v_s=0$. For every value
of $U$ and $H<H_0(U)$ there is a critical value of $n$ at which the system crosses in phase IV and $v_s$
becomes nonzero. The charge velocity is continuous at this critical value of density but the derivative
is discontinuous. Numerical results for the velocities in the presence of a magnetic field are presented
in panels (c)--(f) of Fig.~\ref{velocities}. For $H=0.3$ and $U=\{1,2,4,8\}$ the transition from phase II
to IV can be seen clearly in the discontinuity of the derivative of $v_c$ which coincides with the value
of $n$ for which $v_s$ becomes nonzero (see panels (c) and (d) of Fig.~\ref{velocities}). For $H=0.5$ and
$U=8$ we have $H>H_0(8)=0.4721\ldots$  and for these values of the magnetic field and interaction strength
the system is found in phase II for all densities  which means that $v_c=2 \sin (\pi n)$ and $v_s=0$ for
$n\in[0,1]$. This can be seen in panels (e) and (f) of Fig.~\ref{velocities}.

Fig.~\ref{velocitiesh}  presents the dependence of the spin and charge velocities on magnetic field at
fixed density. Here at low values of $H$ the system is found in phase IV and as the magnetic field is
increased we cross into phase II for which $v_c=2 \sin (\pi n)$ and $v_s=0$.

\begin{figure}
\includegraphics[width=\linewidth]{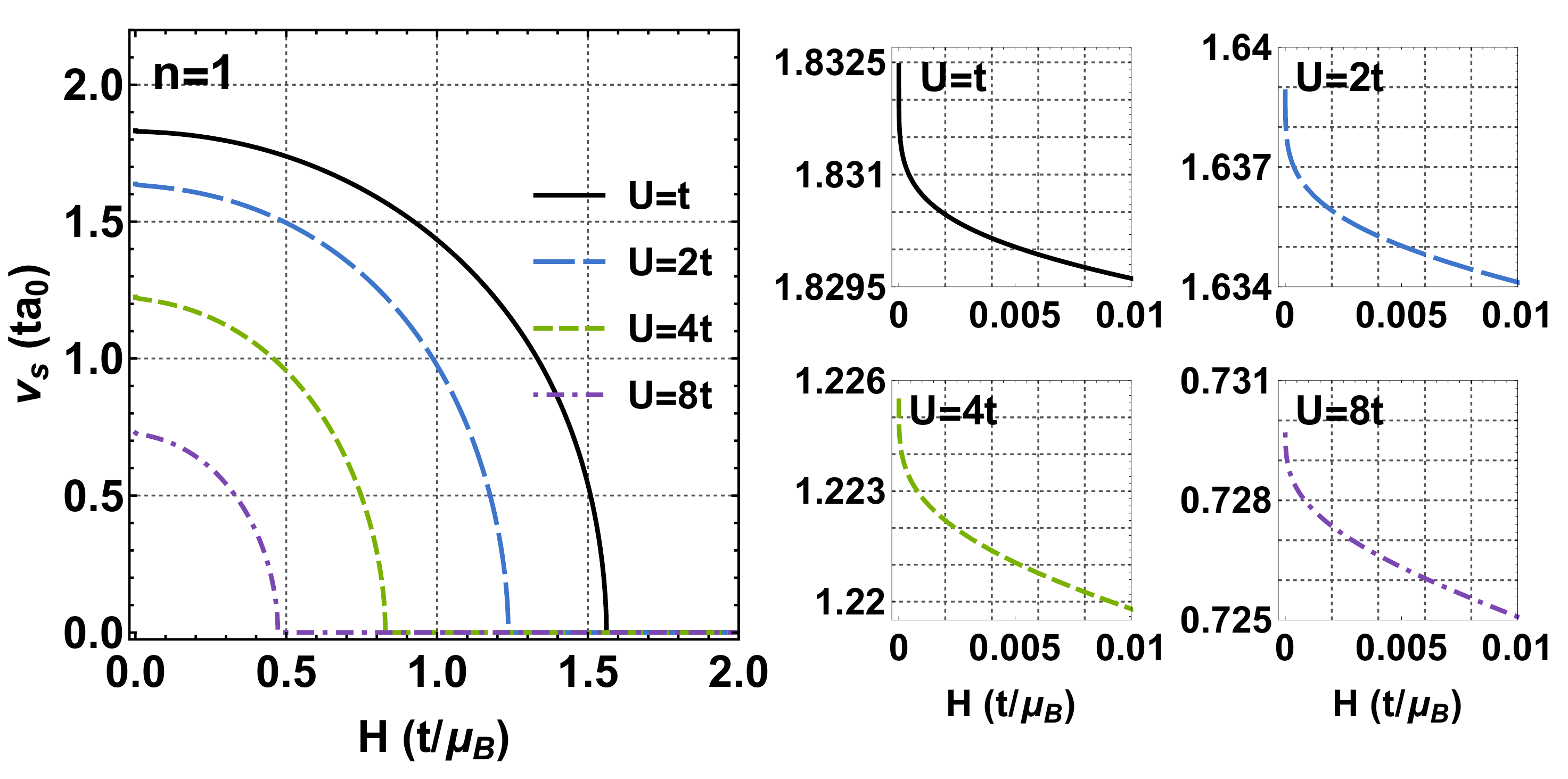}
\caption{(Left panel) The spin velocity at half filling as a function of the magnetic field for
different values of $U$. In the vicinity of $H_0(U)$ the velocity behaves like $v_s\sim (H_0(U)-H)^{1/2}$.
(Right panels) The logarithmic behavior of the spin velocity at small values of $H$. }
\label{velocitieshf}
\end{figure}

At half filling only the spin degrees of freedom are gapless and $v_c=0$. The dependence of the spin
velocity at half filling as a function of the magnetic field is presented in Fig.~\ref{velocitieshf}.
$v_s$ is monotonically decreasing as $H$ is increased and vanishes like $(H_0(U)-H)^{1/2}$ in the vicinity
of $H_0(u)=2\left(1+u^2\right)^{1/2}-2 u$. Another interesting feature is the logarithmic behaviour
$v_s(H)\sim v_s(0)+a/\log(H/H_0)$ for small values of $H$ as it can be seen in the right panels of
Fig.~\ref{velocitieshf}.

\subsection{Susceptibilities at zero temperature}\label{dressedcharge}

The spin and charge susceptibilities in the grand-canonical
  ensemble are defined as
\be
\chi(\mu,H)=2\, \frac{\6 m}{\6 H}\, , \ \ \  \kappa(\mu,H)=\frac{\6 n}{\6 \mu}\, .
\ee
Below we list known analytical formulae for the susceptibilities in each of the five phases at zero
temperature. Our presentation follows Chap.~VI of \cite{EFGKK} (we should point out that our spin
susceptibility contains an additional factor of 2 compared with the definition employed in \cite{EFGKK}).

\begin{itemize}[leftmargin=*]
\item{\bf Phases I and III.}
In these phases the spin and charge susceptibilities are both zero.

\item {\bf Phase II.} The system is fully polarized with density $n=\arccos\left(1-\frac{\mu-\mu_0(H)}{2}\right)/\pi$
and
\be\label{susc2}
\kappa(\mu,H)=\frac{1}{\pi v_c}=\frac{1}{\pi\left(4-(\mu+2u+H)^2\right)^{1/2}}\, .
\ee
Note that $\kappa(\mu,H)$ and the grand-canonical spin susceptibility
$\chi(\mu,H)$ are identical, implying that the canonical spin susceptibility
is zero.

\item{\bf Phase IV.} We introduce an important quantity called the dressed charge matrix
and defined by (\cite{Woyn0,IKR,FY,FK1,FK2} and Chap.~VIII of \cite{EFGKK})
\be
Z=\left(\begin{array}{lr}  Z_{cc}&Z_{cs}\\
                           Z_{sc}&Z_{ss}
        \end{array}
\right)
=\left(\begin{array}{lr}  \xi_{cc}(Q)&\xi_{cs}(A)\\
                           \xi_{sc}(Q)&\xi_{ss}(A)
        \end{array}
\right)\, ,
\ee
where $\xi_{ab}(k)\, ,\  a,b\in\{c,s\}$ satisfy the system of integral equations
\begin{subequations}
\begin{align}
\xi_{cc}(k)&=1+\int_{-A}^A d\lambda'\, \xi_{cs}(\lambda') a_1(\lambda'-\sin k)\, ,\\
\xi_{cs}(\lambda)&=\int_{-Q}^Qdk\, \cos k' \xi_{cc}(k')a_1(\sin k' -\lambda)\nonumber\\
&\ \ \ \ \ \ \ \ \ \ \ \ \ \ \ \ -\int_{-A}^A d\lambda' \xi_{cs}(\lambda')a_2(\lambda'-\lambda)\, ,\\
\xi_{sc}(k)&=\int_{-A}^A d\lambda' \xi_{ss}(\lambda')a_1(\lambda'-\sin k)\, ,\\
\xi_{ss}(\lambda)&=1+\int_{-Q}^Q dk' \cos k'\xi_{sc}(k')a_1(\sin k'-\lambda)\nonumber\\
&\ \ \ \ \ \ \ \ \ \ \ \ \ \ \ \ -\int_{-A}^A d\lambda'\xi_{ss}(\lambda')a_2(\lambda'-\lambda)\, .
\end{align}
\end{subequations}
The susceptibilities are expressed in terms of the elements of dressed charge matrix and
velocities as
\begin{align}
\chi(\mu,H)&=\frac{(Z_{cs}-2Z_{ss})^2}{2\pi v_s}+\frac{(Z_{cc}-2Z_{sc})^2}{2\pi v_c}\, ,\label{susc4}\\
\kappa(\mu,H)&=\frac{Z_{cc}^2}{\pi v_c}+\frac{Z_{cs}^2}{\pi v_s}\, .\label{comp4}
\end{align}

\item{\bf Phase V.} At half filling and magnetization $m<1/2$ we have
\be\label{susc5}
\chi(H,n=1)=\frac{4 \xi_s^2(A)}{\pi v_s}\, , \ \ \ \kappa(H,n=1)=0\, ,
\ee
where $\xi_s(A)$ is the dressed charge for the half filled band and satisfies the
integral equation
\be
\xi_s(\lambda)=1-\int_{-A}^A d\lambda'\, a_2(\lambda-\lambda')\xi_s(\lambda')\, .
\ee
\end{itemize}

\section{Thermodynamics at finite temperature}\label{s3}

The first thermodynamic description of the Hubbard model was derived by Takahashi in the framework of TBA and
assuming the string hypothesis \cite{Tak6}. While extremely important the infinite system of non-linear integral
equations derived using this method is very hard to implement numerically. For this reason in this paper we are
going to investigate the thermodynamic properties of the Hubbard model using the quantum transfer matrix
formalism \cite{Suz,SI,Koma,SAW,K1,K2} which has the advantage of providing a thermodynamic description involving
a finite number of auxiliary functions \cite{JKS}. More precisely we employ only  6 auxiliary functions denoted by
$\mathfrak{b}^\pm\, , \mathfrak{c}^\pm $ and $\overline{\mathfrak{c}^\pm}\, ,$  and we define
\begin{eqnarray*}
&\mathfrak{B}^\pm:=1+\mathfrak{b}^\pm\, ,&\ \ \ \overline{\mathfrak{B}^\pm}:=1+1/\mathfrak{b}^\pm\, ,\\
&\mathfrak{C}^\pm:=1+\mathfrak{C}^\pm\, ,&\ \ \ \overline{\mathfrak{C}^\pm}:=1+\overline{\mathfrak{c}^\pm}\, ,\\
&\Delta \ln \mathfrak{C}:=\ln(\mathfrak{C}^+ / \mathfrak{C}^-)\, ,\ \ \ \  &\Delta \ln \overline{\mathfrak{C}}
:=\ln(\overline{\mathfrak{C}^+} / \overline{\mathfrak{C}^-})\, .
\end{eqnarray*}
These  functions satisfy the following system of nonlinear integral equations derived in  \cite{JKS}
\begin{subequations}\label{qtm}
\begin{align}
\ln \mathfrak{b}^\pm &=-\beta H -K_{2,\pm \alpha-\alpha}*\ln\mathfrak{B}^++K_{2,\pm \alpha+\alpha}*\ln \mathfrak{B}^- \nonumber\\
&\ \ \ \ \ \ \ \ \ \ \ \ \ \ \ \ \ \ \ \ \   -\overline{K}_{1,\pm\alpha}\bullet \Delta\ln(\overline{\mathfrak{c}}/\overline{\mathfrak{C}})\, ,\\
\ln \mathfrak{c}^\pm &=\Psi_c^{\pm} +\overline{K}_{1,-\alpha}*\ln\overline{\mathfrak{B}^+} - \overline{K}_{1,\alpha}*\ln \overline{\mathfrak{B}^-}\nonumber\\
&\ \ \ \ \ \ \ \ \ \ \ \ \ \ \ \ \ \ \ \ \   +\overline{K}_{1,0}\bullet \Delta\ln \overline{\mathfrak{C}}\pm \frac{1}{2}\Delta\ln \overline{\mathfrak{C}}\, ,\\
\ln \overline{\mathfrak{c}^\pm}&=\overline{\Psi}_c^{\pm} -K_{1,-\alpha}*\ln\mathfrak{B}^+ + K_{1,\alpha}*\ln \mathfrak{B}^-\nonumber\\
&\ \ \ \ \ \ \ \ \ \ \ \ \ \ \ \ \ \ \ \ \  - K_{1,0}\bullet \Delta\ln \mathfrak{C}\pm \frac{1}{2}\Delta\ln \mathfrak{C}\, ,
\end{align}
\end{subequations}
where $u=U/4\, ,\ 0<\alpha<u\, , f_\alpha(x)=f(x+i\alpha)$ and
\begin{subequations}\label{kernels}
\begin{align}
K_1(x)&=\frac{u/\pi}{x(x+2 i u)}\, ,\\
\overline{K}_1(x)&=\frac{u/\pi}{x(x-2 i u)}\, ,\\
K_2(x)&=\frac{2 u/\pi}{x^2+4u^2}\, .
\end{align}
\end{subequations}
The driving terms are
\begin{align}
\Psi_c^\pm(x)&=-\beta U/2 +\beta(\mu+H)+\ln \phi_{\pm 0}(x)\, , \\
\overline{\Psi}_c^\pm(x)&=-\beta U/2 -\beta(\mu+H)-\ln \phi_{\pm 0}(x)\, ,\\
\ln \phi_{\pm 0}(x)&=\pm 2\beta(1-x^2)^{1/2}\, ,
\end{align}
and the two types of convolutions appearing in Eqs.~(\ref{qtm}) are defined by
\begin{align}
K* f&=\inti K(x-y) f(y)\, dy\, , \\
K\bullet f&=\mbox{p.v.}\int_{-1}^{+1} K(x-y) f(y)\, dy\, ,
\end{align}
where $\mbox{p.v.}$ denotes the principal value integral. The grandcanonical potential of the
system can be obtained from
\begin{align*}
-&\beta \phi(\mu,H,T,U)=\beta(\mu+H/2+U/4)\nonumber\\
&  -\int_{-1}^{+1} \mathcal{K}\ln[(1+\mathfrak{c}^++\overline{\mathfrak{c}^+})(1+\mathfrak{c}^-+\overline{\mathfrak{c}^-})]\, dx\nonumber\\
&  + \inti[(\mathcal{K}_{\alpha-2u}-\mathcal{K}_\alpha)\ln \mathfrak{B}^+ \nonumber\\
&\ \ \ \ \ \ \ \ \ \ \ \ \ \ \ \ \ \ \ - (\mathcal{K}_{-\alpha-2u}-\mathcal{K}_{-\alpha})\ln \mathfrak{B}^-]\, dx
\end{align*}
with  $\mathcal{K}(x)=-[2\pi (1-x^2)^{1/2}]^{-1}$.
In the noninteracting case, $U=0$, the grandcanonical potential is known analytically
\begin{align}
&\phi_{FF}(\mu,H,T)=-\frac{T}{2\pi}\int_{-\pi}^{\pi} dk\, \ln \left[1+e^{\frac{2\cos k -\mu-H}{T}}\right]\nonumber \\
&\ \ \ \ \ \ \ \ \ \ \ \ \  \ \ \ \ \ \  -\frac{T}{2\pi}\int_{-\pi}^{\pi} dk\, \ln \left[1+e^{\frac{2\cos k -\mu+H}{T}}\right]\, ,
\end{align}
and in the limit of infinite repulsion we have \cite{Tak1}
\begin{align}
&\phi_{\infty}(\mu,H,T)=\nonumber\\
&\ \ \ \   -\frac{T}{2\pi}\int_{-\pi}^{\pi} dk\, \ln \left[1+2 \cosh\left(\frac{H}{T}\right)e^{\frac{2\cos k -\mu}{T}}\right]\, .
\end{align}

The integral equations (\ref{qtm}) can be solved by a simple iterative
procedure. First, we make an initial guess of the six functions, which are
then plugged into the right hand side of (\ref{qtm})
obtaining an approximate solution.  This process is iterated until the
difference between the functions obtained in two successive steps is smaller
than a given error. This algorithm requires an efficient numerical treatment
of the two types of convolutions appearing in the integral
equations which is detailed in Appendices \ref{app1} and \ref{app2}. Another
difficulty lies in the fact that while for the first type of convolution we
use the Fast Fourier Transform for the second type we use a Chebyshev
quadrature which means that the six functions are discretized on different
grids requiring the use of interpolation at each iterative step.

\section{Quantum critical behaviour}\label{s4}

At low temperatures the Hubbard model shows thermodynamically
  activated behaviour in the gapped phases I and III, and algebraic dependence
  on temperature in the gapless phases II, IV and V, see (\ref{lowT}). The
  various types of behaviour hold inside the phases. At the boundaries -- here
  referred to as quantum critical lines -- rather complex crossover behaviour
  is observed which is one of the main objectives of this paper.

The Hubbard model presents a multitude of quantum phase transitions and
quantum critical lines.  The effects of the QPTs can be measured at low but
finite temperatures in the quantum critical (QC) region which is characterized
by strong coupling of quantum and thermal fluctuations \cite{Sachdev}.  In the
vicinity of the quantum critical points the thermodynamics is expected to be
universal and determined by the universality class of the transition. For
example, in the case of a QPT induced by the variation of the chemical
potential at fixed magnetic field the pressure is assumed to satisfy \cite{ZH}
\be\label{scaling}
p(\mu,H,T)\sim p_r(\mu,H,T)+T^{\frac{d}{z}+1}\mathcal{P}_H\left(\frac{\mu-\mu_c(H)}
{T^{\frac{1}{\nu z}}}\right)\, ,
\ee
with $p_r$ the regular part of the pressure, $d$ the dimension, $\mathcal{P}_H$ a universal
function and $\mu_c(H)$ the quantum critical point. The correlation length exponent $\nu$ and
the dynamical critical exponent $z$ determine the universality class of the transition. Other
relevant thermodynamic quantities can be derived from (\ref{scaling}) using  thermodynamic
identities.

The universality class of a QPT can be determined by plotting the ``scaled" quantity $(p(\mu,H,T)-
p_r(\mu,H,T))T^{-\frac{d}{z}-1}$ (or other suitable thermodynamic function)  as a
function of the chemical potential and for several  values of temperature \cite{ZH}.
If the $z$ and $\nu$ exponents are chosen correctly then all the curves intersect at $\mu_c(H).$
It is efficient to choose a thermodynamic parameter for which the regular part is
known from previous theoretical considerations. Plotting the ``scaled" quantities as
functions of $(\mu-\mu_c(H))/T^{\frac{1}{\nu z}} $ all curves collapse to the universal
function $\mathcal{P}_H$.

\subsection{QPTs induced by the variation of the chemical potential at zero magnetic field}

\begin{figure}
\includegraphics[width=\linewidth]{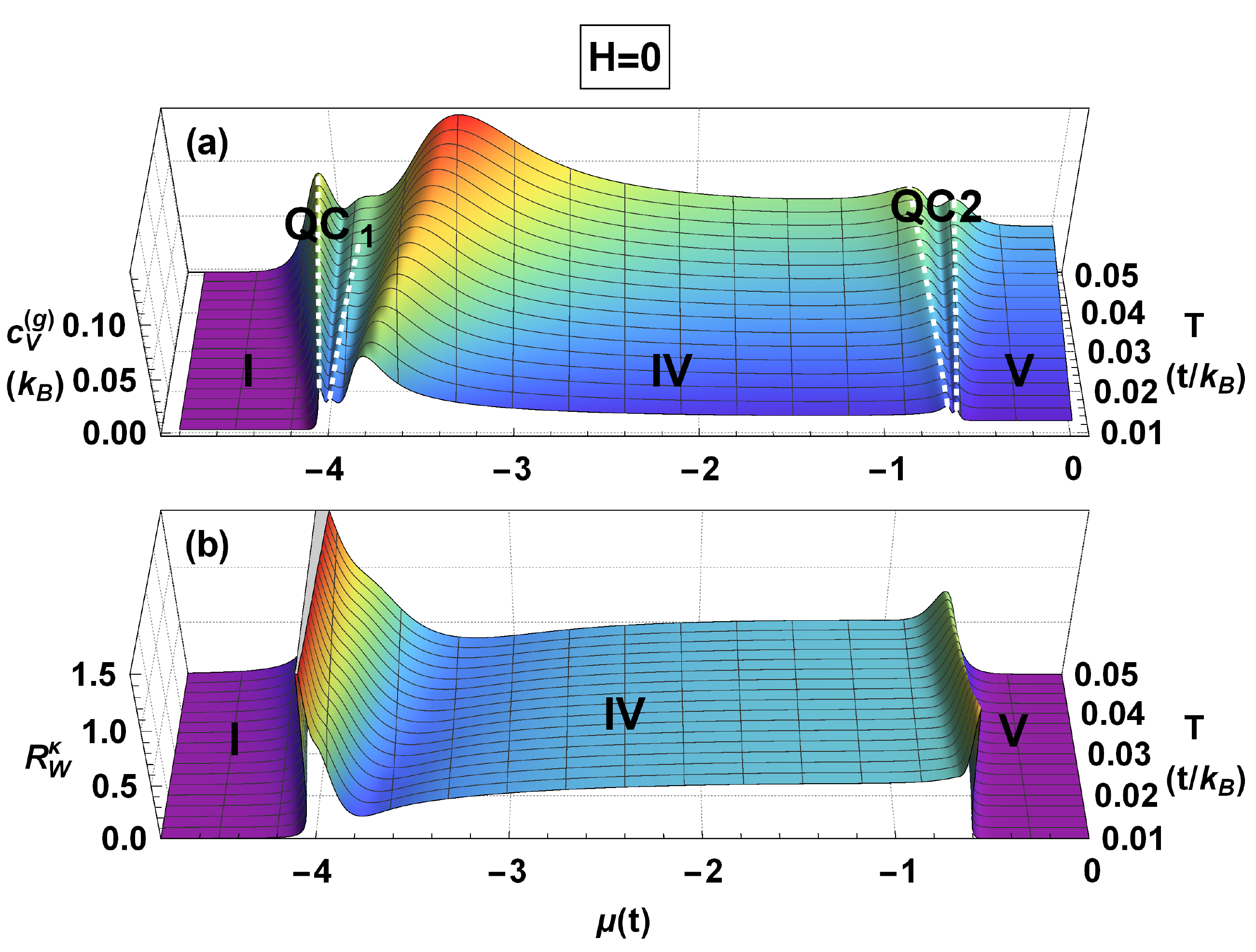}
\caption{(a) Chemical potential and temperature dependence of the grandcanonical specific heat for $U=4\,t$ and
zero magnetic field. The boundaries of the critical regions identified with the maxima of $c_V^{(g)}$ are highlighted
by dashed white lines. The two QCPs are $\mu_c^{(I\rightarrow IV)}=-4\,t$  and $\mu_c^{(IV\rightarrow V)}\equiv \mu_-(0)=-0.6433\,t\, .$
(c) Chemical potential and temperature dependence of the Wilson ratio $R_W^\kappa$. Note the anomalous enhancement
in the quantum critical regions.
}
\label{3D0}
\end{figure}

\begin{figure}
\includegraphics[width=\linewidth]{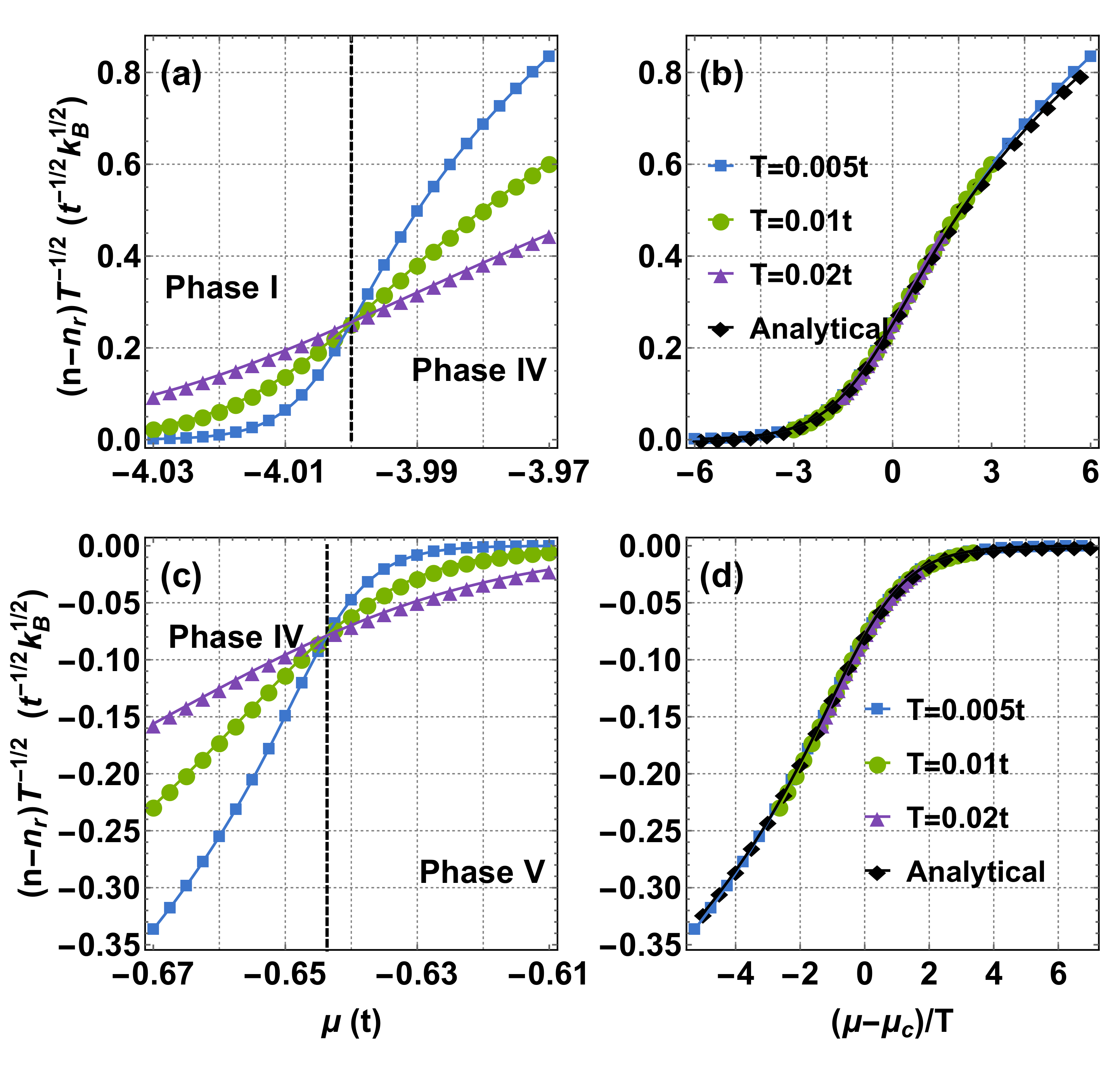}
\caption{Plots of the scaled densities $(n-n_r)T^{-1/2}$ as functions of the chemical potential at different
temperatures in the vicinities of the QCPs: $\mu_c^{(I\rightarrow IV)}=-4\,t$ (a) and $\mu_c^{(IV\rightarrow V)}=-0.6433\,t$ (c). For the
first transition $n_r=0$  and for the second $n_r=1$. The dashed vertical lines pass through the QCPs.
Plotting the scaled densities as  functions of $(\mu-\mu_c)/T$ all curves collapse to the universal function
$\mathcal{P}_H'(x)$ (b) for the first transition and (d)  for the second transition. In panels (b) and (d) the black diamonds
represent the analytical predictions for the  universal function $\mathcal{P}_H'(x)$ obtained from the
derivative of Eq.~(\ref{ph}) (b) and Eq.~(\ref{general}) (d) with $a=0.462$ and $b=-1$.
}\label{scaling0}
\end{figure}

We will first investigate the QPTs induced by the variation of the chemical potential at
zero magnetic field. In the attractive case a similar analysis can be found in \cite{CYBG0,CYBG}.
In addition to the determination of the QCPs and the universality class
of the transitions a problem of considerable importance is represented by the identification
of the boundaries of the QC regions. It was argued recently \cite{HJYLG,YCZSG,BGKRT,PKF,PK}
that the grandcanonical specific  heat $c_V^{(g)}=-T \left(\6^2\phi/\6 T^2\right)_{\mu,H}$
can be used for this task. The specific heat  presents two lines of local maxima fanning out
from the QCP and the location of these maxima can be identified with the boundaries of the
QC region. Another useful quantity which can be  used to distinguish the various phases at
low temperatures \cite{YCLRG} is the compressibility Wilson  ratio defined by
\be
R_W^\kappa=\frac{\pi^2}{3}T \frac{\kappa}{c_V^{(g)}}\, .
\ee
where $\kappa$ is the compressibility of the system.
In Fig.~\ref{3D0} we present the dependence of the grandcanonical specific heat and compressibility
Wilson ratio on chemical potential and temperature for $U=4$ and $H=0$. The system presents
two QPTs between phases I and IV  with  critical point $\mu_c^{(I\rightarrow IV)}=-2u-2=-4$
and between phases IV and V  with  $\mu_c^{(IV\rightarrow V)}\equiv \mu_-(0)=-0.6433\, $ ($\mu_-(0)$
is given by Eq.~(\ref{muminus})). The boundaries of both critical regions can be identified with
the lines of maxima of the specific heat fanning out from the critical  points. The Wilson ratio
presents anomalous enhancements in the QC regions and is almost constant in the other regimes.

The identification of the universality classes of the QPTs is done by employing the scaling
relation for the density which at zero temperature  is zero in phase I and one in phase V.
Taking the derivative with respect to $\mu$  of Eq.~(\ref{scaling}) we obtain
\begin{align*}
n(\mu,H,T)\sim \frac{\6 p_r}{\6 \mu}(\mu,H,T)+T^{\frac{d}{z}+1-\frac{1}{\nu z}}\mathcal{P}_H'
\left(\frac{\mu-\mu_c(H)}{T^{\frac{1}{\nu z}}}\right)\, .\\
\end{align*}
The curves $(n-n_r)T^{-\frac{d}{z}-1+\frac{1}{\nu z}}$ with $n_r=\6 p_r/\6 \mu=0$ for the
first transition and $n_r=1$ for the second transition intersect at the critical points
$\mu_c$ when the critical exponents are $z=2$ and $\nu=1/2$ as it can be seen in
Fig.~\ref{scaling0}.

Even though both QPTs are characterized by the same critical exponents the universal thermodynamics
in the vicinity of the QCPs is not given by the same universal function $\mathcal{P}_H(x).$ For the
transition between phases I to IV we can analytically derive the universal function as follows. Close
to the first critical point the system is characterized by very low densities and it is equivalent to
the repulsive Gaudin-Yang model (two component fermions with repulsive delta-interaction). For this
continuum model the QPT between the vacuum and the TLL phase  belongs to the universality class of
spin-degenerate impenetrable particle gas \cite{PKF} with the universal thermodynamics  described by
Takahashi's formula \cite{Tak1}($x=[\mu-\mu_c(H)]/T\, ,y=H/T$)
\be
\label{tt1}
p=\frac{T^{3/2}}{2\pi}\inti\ln\left[1+(1+e^{-2|y|})
e^{-k^2+x}\right]\, dk\, .
\ee
For $H=0$ we have $e^{-2|y|}=1$ and
\be
\label{ph}
\mathcal{P}_H(x)=\frac{1}{2\pi}\inti\ln\left[1+2
e^{-k^2+x}\right]\, dk\, .
\ee
A comparison of the numerical data with the analytical predictions of Eq.~(\ref{ph}) can be seen
in Fig.~\ref{scaling0} (b) which confirms the validity of our analytical derivation.
 Eq.~(\ref{ph}) is valid for all $U>0$. In the case of free fermions on the
lattice the system still presents a QPT between the vacuum and the partially filled and magnetized
band phase with the QCP $\mu_c(H)=-2-H$ (see Chap.~6.1 of \cite{EFGKK}) with the thermodynamics described
by the free fermionic formula.

We should point out that the QC regions have physical properties which are different from the ones
of the surrounding phases, particularly in the case of the correlation functions. For example, at
low temperatures, phase I can be well described by a dilute classical gas in which thermal effects
play an important role while in phase IV the quantum effects dominate and the system is described by
a two-component TLL. In the QC region between phases I and IV both quantum and thermal effects are
important producing distinct correlation functions.

As we will show below all the other QPTs investigated in this paper have critical exponents
$z=2$ and $\nu=1/2$ and therefore it is sensible to assume that the scaling of the pressure for these
transitions is given  by $p=T^{3/2}\mathcal{P}(x)$ with the universal function
\be\label{general}
\mathcal{P}(x)=\frac{a}{2\pi}\inti\ln\left[1+e^{-k^2+b\, x}\right]\, dk\, ,
\ee
where $a$ and $b$ are free parameters and $x=(\mu-\mu_c(H))/T$ for the
chemical potential induced QPTs and $x=(H-H_c(\mu))/T$ for the magnetic field
ones.
The parameter $a$ takes some positive real number and
  is related to the square root of the mass of the particle with parabolic
  energy-momentum dispersion. The parameter $b$ describes the strength of the
  coupling of the external field, chemical potential or magnetic field, to
  some particle number and typically takes values $\pm 1$ for particle and
  hole type excitations and $\pm 2$ for magnetic excitations: a change of the
  particle number carries energy $\pm\mu$ and a spin-flip results in $\pm 2H$,
  see (\ref{hamc}c). An exception is realized by the phase transition from II
  to IV where particles with parabolic dispersion enter the system on a
  background of majority particles with finite density. There the Hubbard
  interaction leads to an effective parameter $b$ different from -1.

For the transition between phases IV and V the best fit is
  obtained for $a=0.462$ and $b=-1$ and is shown in panel (d) of
  Fig.~\ref{scaling0}.


\subsection{QPTs induced by the variation of the chemical potential at nonzero magnetic field}

\begin{figure}
\includegraphics[width=\linewidth]{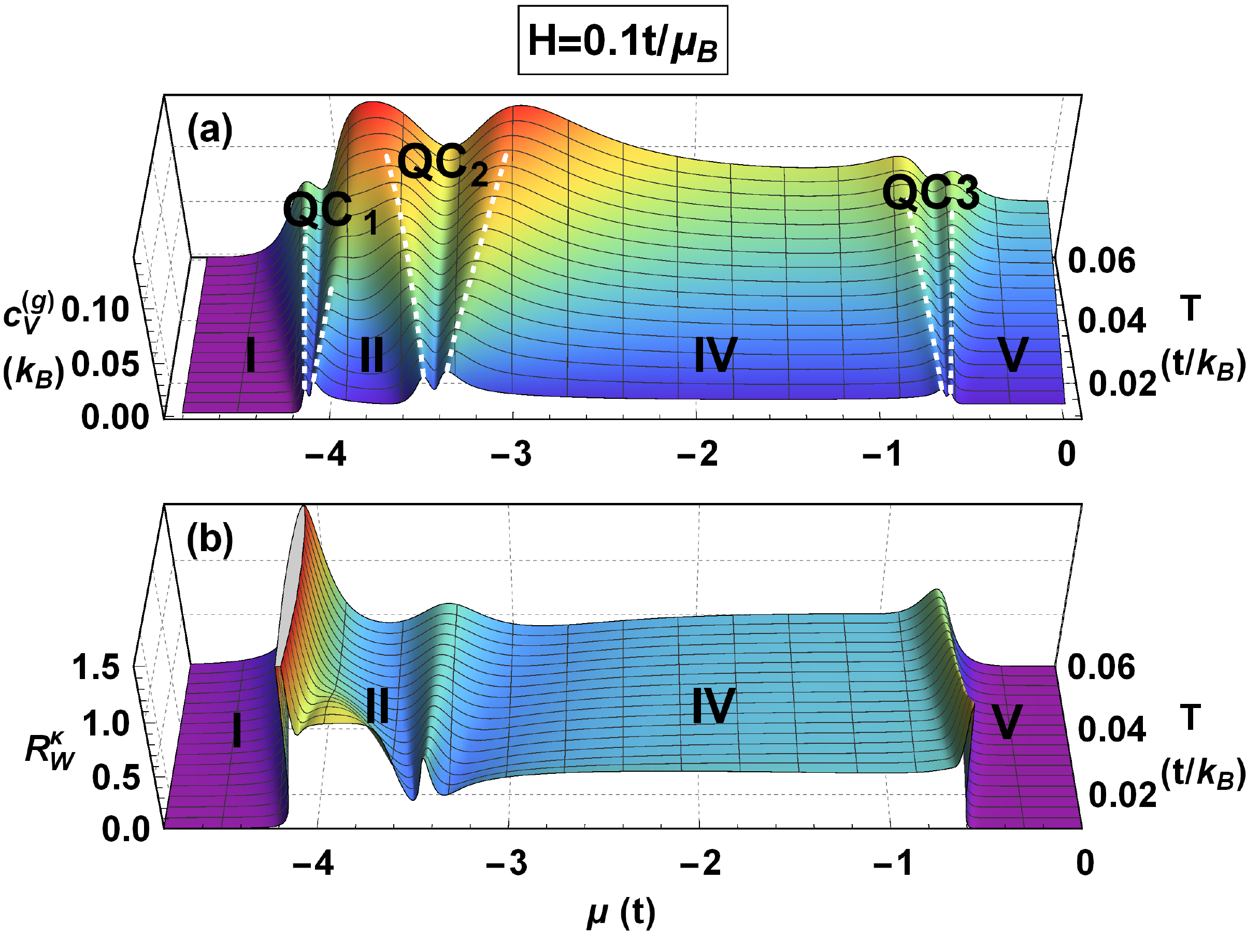}
\caption{(a) Chemical potential and temperature dependence of the grandcanonical specific heat  for $U=4t$ and
$H=0.1\, t/\mu_B$. The boundaries of the critical regions identified with the maxima of $c_V^{(g)}$ are highlighted
by dashed white lines. The three QCPs are $\mu_c^{(I\rightarrow II)}=-2u-2-H=-4.1\,t\, ,$
$\mu_c^{(II\rightarrow IV)}=-3.429\,t\, $ and $\mu_c^{(IV\rightarrow V)}=-0.6453\,t$.
(b) Chemical potential and temperature dependence of the Wilson ratio $R_W^\kappa$.
}
\label{3D1}
\end{figure}

In the presence of a magnetic field the repulsive Hubbard model presents three QPTs induced by
the variation of the chemical potential. It can be seen in Fig.~\ref{3D1} for $U=4$ and $H=0.1$ that
the specific heat presents two lines of maxima to the left and right of each  QCP and that the
Wilson ratio also presents  maxima in each quantum critical region while being almost constant
outside of them. The values of the three QCPs are: $\mu_c^{(I\rightarrow II)}=-2u-2-H=-4.1\, ,$
$\mu_c^{(II\rightarrow IV)}=-3.429\, $ and $\mu_c^{(IV\rightarrow V)}=-0.6453$.

For the determination of the critical exponents we use the scaling of the densities for the
I $\rightarrow$ II and  IV $\rightarrow$ V transitions ($n_r$(phase I)=0 and $n_r$(phase V)=1)
and the density of down spins for the II $\rightarrow$ IV transition ($n_{\downarrow,r}$(phase II)=0).
Using $n_\downarrow=-(\6 \phi/\6 \mu+\6 \phi/\6 H)/2$ we find
\begin{figure}
\includegraphics[width=\linewidth]{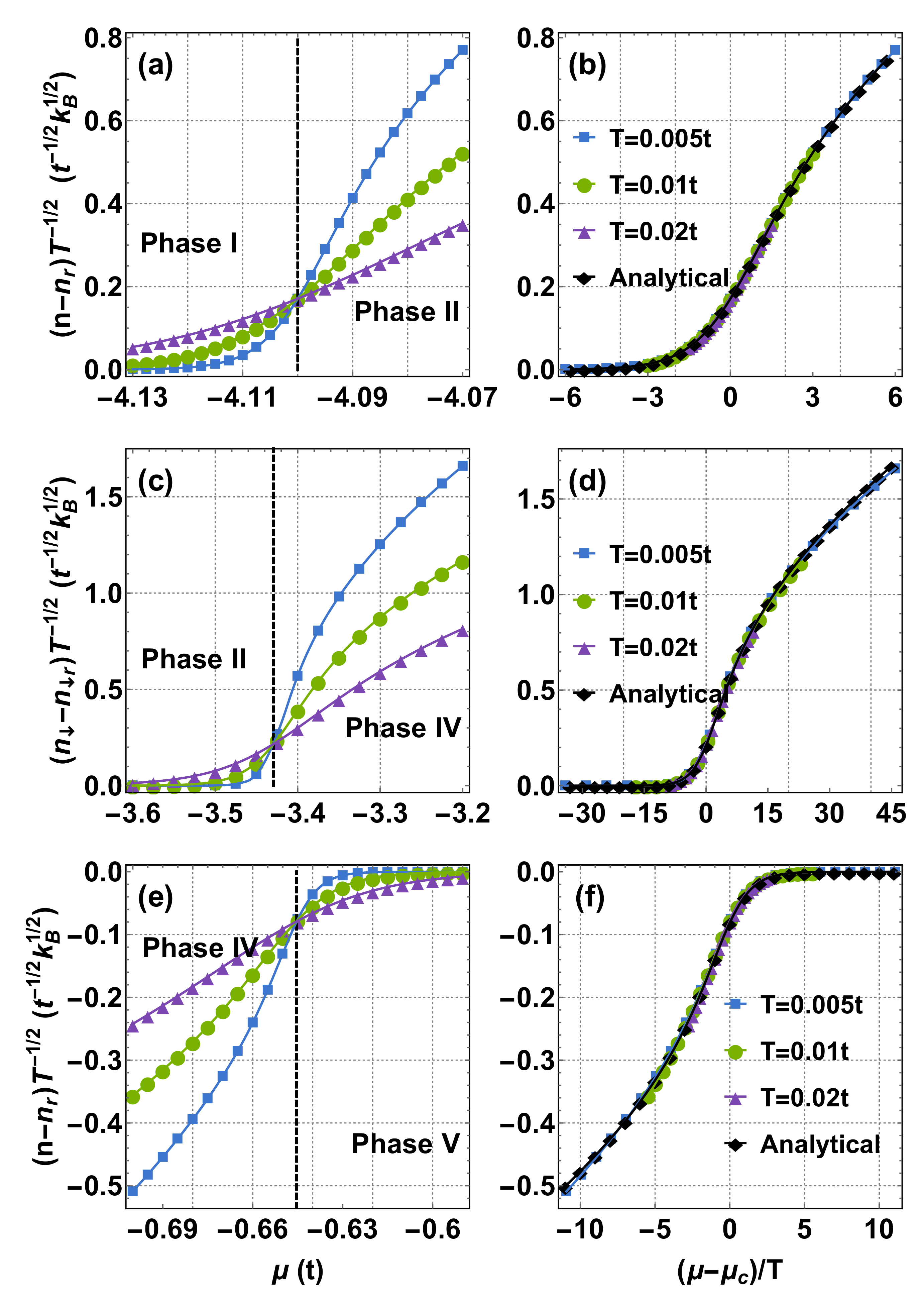}
\caption{Plots of the scaled total densities $(n-n_r)T^{-1/2}$ and density of down spins
$(n_\downarrow-n_{\downarrow r})T^{-1/2}$   as functions of the chemical potential at
different temperatures in the vicinities of the QCPs $\mu_c^{(I\rightarrow II)}=-4.1\,t$ (a),
$\mu_c^{(II\rightarrow IV)}=-3.429\,t\, $ (c) and $\mu_c^{(IV\rightarrow V)}=-0.6453\,t$ (e).
The regular parts are $n_r$(phase I)=0, $n_{\downarrow,r}$(phase II)=0 and $n_r$(phase V)=1.
The dashed vertical lines pass through the QCPs. Plotting the scaled quantities as  functions
of $(\mu-\mu_c)/T$ all curves collapse to the universal functions $\mathcal{P}_H'(x)$ (b), (d)
and (f). In panel (b) the black diamonds represent the analytical predictions for the
universal function $\mathcal{P}_H'(x)$ obtained from the derivative of Eq.~(\ref{ph1}).
In panels (d) and (f) the analytical predictions for the universal functions are obtained from
Eq.~(\ref{general}) with $a=3.1\, , b=0.4$ (d) and $a=0.476\, ,b=-1$ (f).
}\label{scaling1}
\end{figure}

\begin{align*}
&n_\downarrow(\mu,H,T)\sim n_{\downarrow,r}(\mu,H,T) \\
&\ \ \ \ \ \ \ \ \ \ +\frac{(1-\mu_c'(H))}{2}T^{\frac{d}{z}+1-\frac{1}{\nu z}}\mathcal{P}_H'
\left(\frac{\mu-\mu_c(H)}{T^{\frac{1}{\nu z}}}\right)\, .\\
\end{align*}

For all QPTs the critical exponents are $z=2$ and $\nu=1/2$ as it can be seen from
Fig.~\ref{scaling1}. Similar to the previous case, even though the critical exponents
are the same the universal function $\mathcal{P}_H(x)$ seems to be different for
each QPT. For the I $\rightarrow$ II transition from  Eq.~(\ref{tt1}) ($e^{-2|y|}\sim 0$
at low temperatures) the universal function at finite magnetic field is
\be
\label{ph1}
\mathcal{P}_H(x)=\frac{1}{2\pi}\inti\ln\left[1+ e^{-k^2+x}\right]\, dk\, .
\ee In panel (b) of Fig.~\ref{scaling1} it is shown that this analytical
formula agrees perfectly with the numerical data. Together
  with the previous result this proves that the transition I $\rightarrow$ IV
  belongs to the universality class of the {\em spin-degenerate impenetrable
    particle gas} which is characterized by Takahashi's formula
  Eq.~(\ref{tt1}). The transition I $\rightarrow$ II belongs to the strong
  field limit of the spin-degenerate impenetrable particle gas and is better
  known as simply the {\em impenetrable particle gas}.

For the II to IV and IV to V transitions the best fits for the universal function are obtained
using Eq.~(\ref{general}) with  $a=3.1\, , b=0.4$ (II $\rightarrow$ IV) and $a=0.476\, ,b=-1$
(IV $\rightarrow$ V) as it can be seen in panels (d) and (f) of Fig.~\ref{scaling1}.

\subsection{QPTs induced by the variation of the magnetic field at half filling}

\begin{figure}
\includegraphics[width=\linewidth]{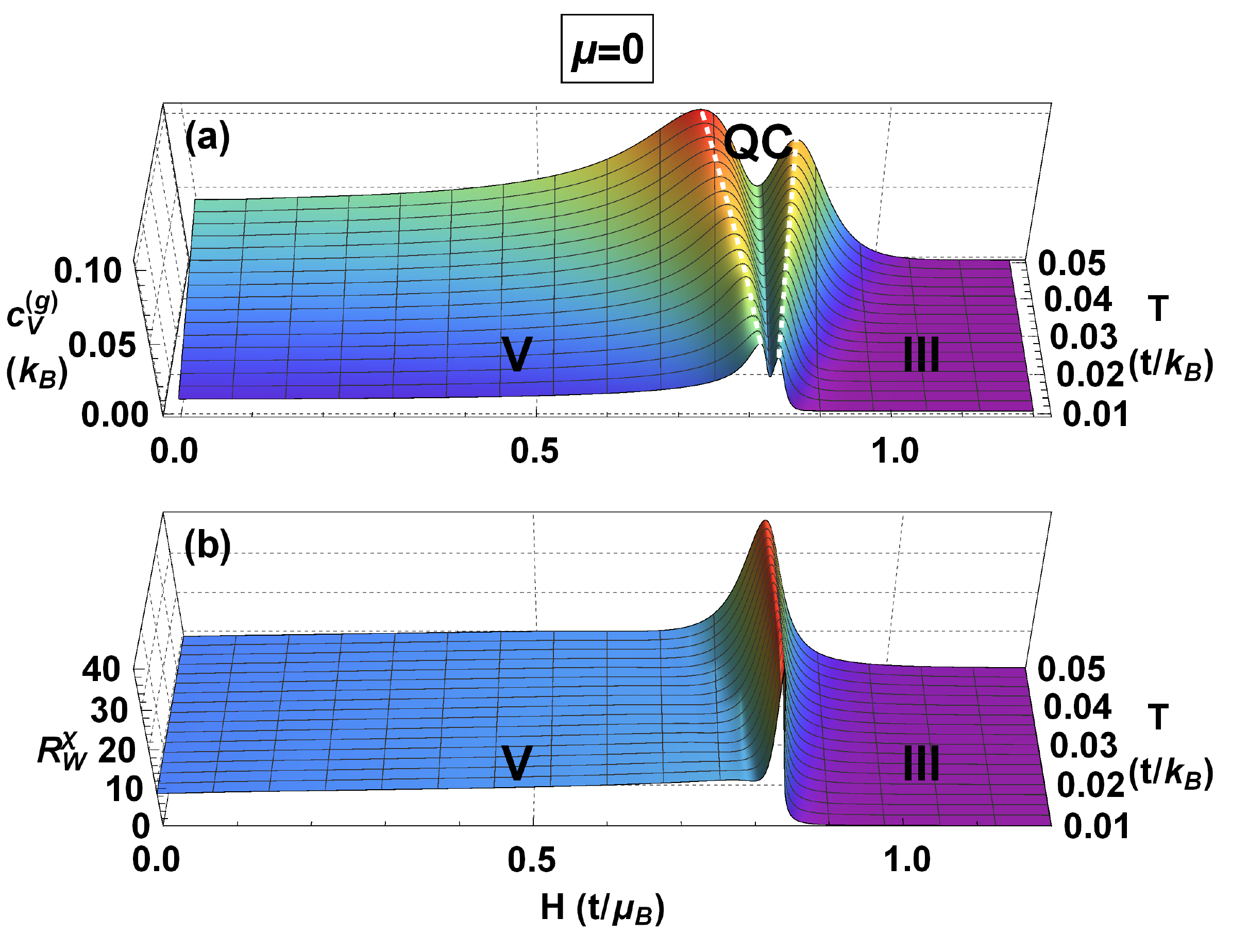}
\caption{(a) Magnetic field and temperature dependence of the grandcanonical specific heat  for $U=4t$
at half filling ($\mu=0$). The boundaries of the critical region identified with the maxima of $c_V^{(g)}$ are highlighted
by dashed white lines. The QCP is $H_c^{(V\rightarrow III)}\equiv H_0(U)= 0.8284\, t/\mu_B\, .$
(b) Magnetic field and temperature dependence of the Wilson ratio $R_W^\chi$.
}
\label{3DHF}
\end{figure}
\begin{figure}
\includegraphics[width=\linewidth]{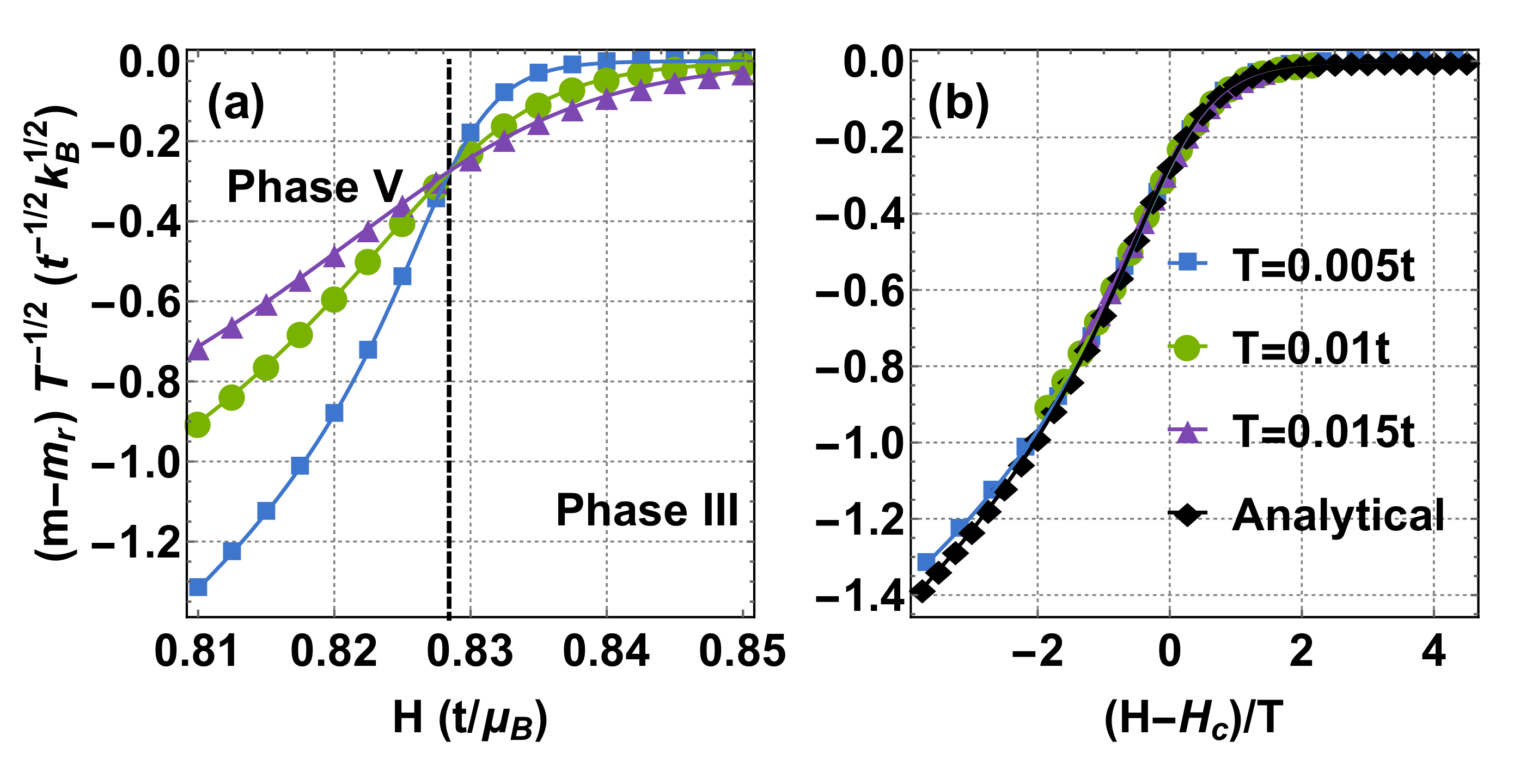}
\caption{ (a) Plot of the scaled magnetization $(m-m_r)T^{-1/2}$ versus magnetic field at
different temperatures in the vicinity of the  QCP $H_c^{(V\rightarrow III)}\equiv H_0(U)= 0.8284\, t/\mu_B\, .$
The regular part of the magnetization is  $m_r$(phase III)=$1/2.$
(b) Universal function $\mathcal{P}_\mu'(x)$  obtained from the collapse of all the curves plotted
as functions of  $(H-H_c)/T$. The black diamonds represent the analytical predictions for the
universal function $\mathcal{P}_{\mu}'(x)$ obtained from the derivative of Eq.~(\ref{general})
for $a=0.8$ and $b=-2$.
}\label{scalingHF}
\end{figure}

The Hubbard model also presents QPTs induced by the variation of the magnetic field when the
chemical potential is fixed. The most interesting is the transition between phases V and III
at half filling. For magnetically induced phase transitions the scaling relation (\ref{scaling})
becomes
\be\label{scalingh}
p(\mu,H,T)\sim p_r(\mu,H,T)+T^{\frac{d}{z}+1}\mathcal{P}_\mu\left(\frac{H-H_c(\mu)}
{T^{\frac{1}{\nu z}}}\right)\, .
\ee
Also in this case the relevant dimensionless ratio is the susceptibility Wilson ratio \cite{GYFBL}
defined by
\be
R_W^\chi=\frac{4}{3}\pi^2 T \frac{\chi}{c_V^{(g)}}\, ,
\ee
where $\chi$ is the magnetic susceptibility.
The dependence on the magnetic field and temperature of the grand canonical specific heat and
Wilson ratio $R_{W}^\chi$ for $U=4$ and $\mu=0$ is presented in Fig.~\ref{3DHF}. The specific heat
presents lines of  maxima which define the boundaries of the CR and the Wilson ratio presents anomalous
enhancement  in the vicinity of the QCP defined by $H_c\equiv H_0(U)= 0.8284$ ($H_0(U)$ is defined
in Eq.~(\ref{H0}). The scaling of the magnetization which satisfies
\begin{align*}
2 m(\mu,H,T)\sim m_r(\mu,H,T)+T^{\frac{d}{z}+1-\frac{1}{\nu z}}\mathcal{P}_\mu '
\left(\frac{H-H_c(\mu)}{T^{\frac{1}{\nu z}}}\right)\, ,\\
\end{align*}
is presented in Fig.~\ref{scalingHF} ($m_r$(phase III)=1/2). The critical exponents of this transition are
also $z=2$ and $\nu=1/2$.

The universal function is given by Eq.~(\ref{general}) with $a=0.8$ and $b=-2$. The value for $b$ agrees
with the one obtained in \cite{CYBG} for the attractive Hubbard model. It should be noted that
the connection between the thermodynamics of the repulsive and attractive model is given by (see Chap.~II of \cite{EFGKK})
\be\label{symmHmu}
\phi(\mu,H,T,u)=\phi(H,\mu,T,-u)-\mu+H\, ,
\ee
and the equivalent transition in the attractive case is a chemical potential induced one (I to V in the
terminology of \cite{CYBG} and their $x$ is $-x$ in Eq.~(\ref{general})).

\section{Thermodynamic properties  below half filling}\label{s5}

\begin{figure}
\includegraphics[width=\linewidth]{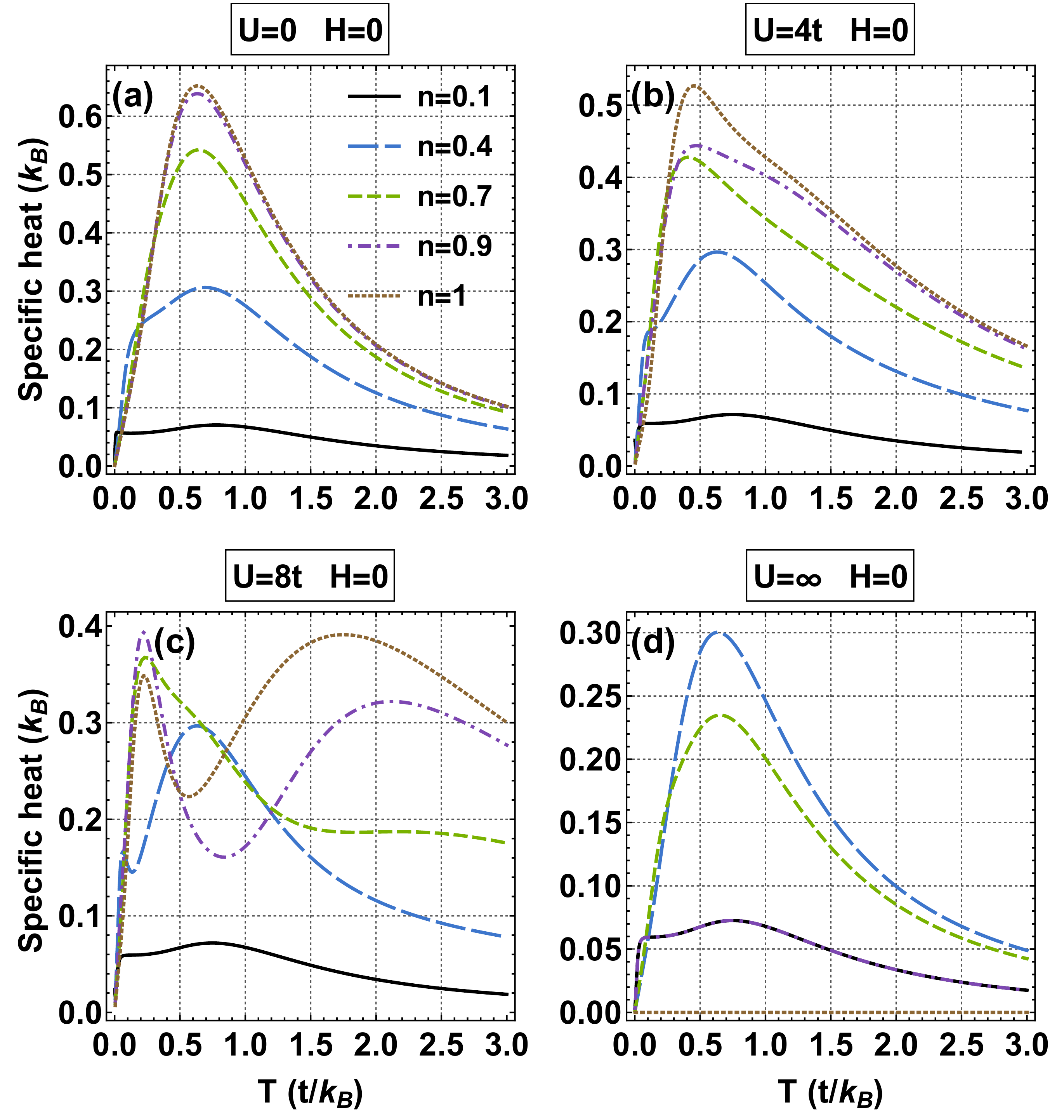}
\caption{Temperature dependence of the specific heat in zero magnetic field  for various filling
fractions and interaction strengths.
}\label{sheatn}
\end{figure}

In this Section we will investigate the effect of the magnetic field on the thermodynamic
properties of the repulsive Hubbard model below half filling, $n\in[0,1)$.
We will mainly focus on the canonical specific heat, the grand canonical
magnetic susceptibility and compressibility defined by
\begin{align}
c_V^{(c)}&=-T\left[\left(\frac{\6^2 \phi}{\6 T^2}\right)_{\mu,H}+\left(\frac{\6 n}
{\6 T}\right)_{\mu,H}^2\left(\frac{\6 n}{\6 \mu}\right)_{T,H}^{-1}\right]\, , \\
\chi&=-\left(\frac{\6^2 \phi}{\6 H^2}\right)_{\mu,T}\, , \\
\kappa&=-\left(\frac{\6^2 \phi}{\6 \mu^2}\right)_{H,T}\, .
\end{align}
The thermodynamic properties of the system with $n\in(1,2]$ are related
to similar quantities below half filling using the following
symmetry of the grand canonical potential (Chap.~II of \cite{EFGKK})
\be\label{symm}
\phi(\mu,H,T,u)+\mu=\phi(-\mu,H,T,u)-\mu\, .
\ee
(Note that this elegant symmetry relation is literally satisfied if we extend
in (\ref{hamc}c) the term $\mu(n_{j,\uparrow}+n_{j,\downarrow})$ to
$\mu(n_{j,\uparrow}+n_{j,\downarrow}-1)$.)
For the densities and magnetization we find $(\mu\ge0\, , H\ge0)$
\begin{subequations}\label{right}
\begin{align}
n(\mu)&=2-n(-\mu)\, ,\ \
m(\mu)=m(-\mu)\, ,\\
n_\uparrow(\mu)&=1-n_\downarrow(-\mu)\, ,\ \
n_\downarrow(\mu)=1-n_\uparrow(-\mu)\, ,
\end{align}
\end{subequations}
which shows that for our Hamiltonian (\ref{ham}) the system is at half filling
for $\mu=0$. From Eq.~(\ref{symm}) the connection between the thermodynamic quantities
below and above half filling is given by $(n\in[0,1))$
\be
c_V^{(c)}(n)=c_V^{(c)}(2-n)\, ,\   \chi(n)=\chi(2-n)\, ,\  \kappa(n)=\kappa(2-n)\, .
\ee

\subsection{Specific heat}

\begin{figure*}
\includegraphics[width=\linewidth]{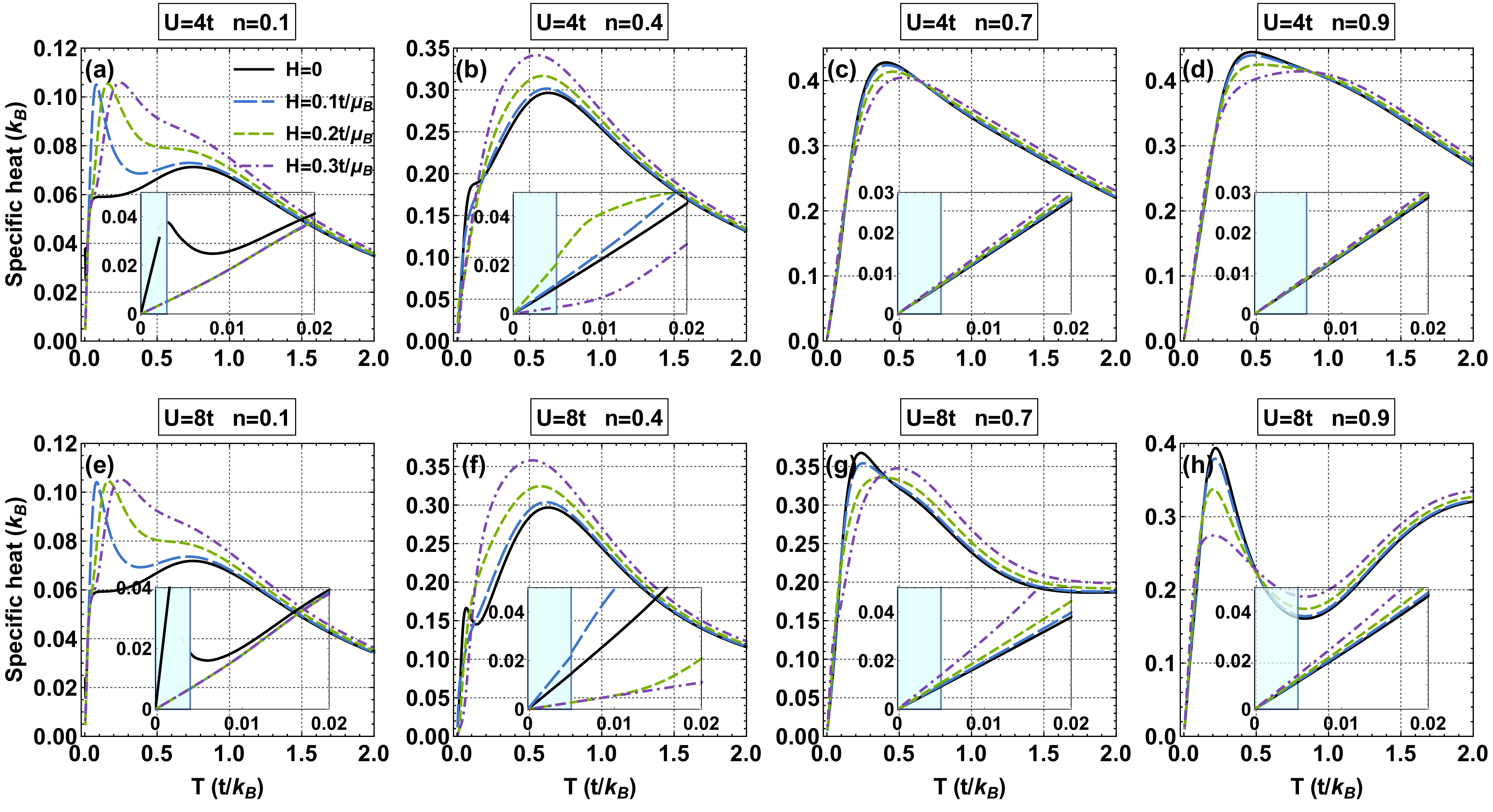}
\caption{Temperature dependence of the specific heat in the presence of a magnetic field. The insets present a zoom of the
data at low temperatures. The shaded regions  contain the theoretical predictions $c_V^{(c)}=\gamma T$
with specific heat coefficients given in Tables \ref{tablesh1} and \ref{tablesh2}.  }
\label{sheath}
\end{figure*}

The first numerical investigations of the specific heat were performed by Shiba and Pincus \cite{Shiba1,Shiba2}
who considered finite chains for up to 6 lattice sites at half filling. The low-temperature properties
for the entire phase diagram were studied by Takahashi \cite{Tak7} using the TBA equations \cite{Tak6}.
For small temperatures the specific heat in phases II, IV and V is linear in $T$ and in phases I and III
it behaves like $T^{3/2} e^{-\alpha/ T}$. On the critical lines $c_V^{(c)}$ is proportional to $T^{1/2}.$
The specific heat coefficient which characterizes the linear dependence on the temperature ($c_V^{(c)}\sim
\gamma\, T$) is given by
\begin{subequations}
\begin{align}
\gamma_{II}&=\frac{\pi }{3}\frac{1}{v_c}\, ,\ \ &\mbox{phase II}\, ,\\
\gamma_{IV}&=\frac{\pi}{3}\left(\frac{1}{v_c}+\frac{1}{v_s}\right)\, ,\ \ &\mbox{phase IV}\, ,\\
\gamma_{V}&=\frac{\pi}{3}\frac{1}{v_s}\, ,\ \ &\mbox{phase V}\, .
\end{align}
\end{subequations}
with $v_{c,s}$ the charge and spin velocities. Numerical investigations of the
specific heat using the TBA equations can be found in \cite{UKO2,KUO1,Ha}. At
zero magnetic field extensive results derived using the QTM thermodynamics
(\ref{qtm}) can be found in \cite{JKS} and Chap.~XIII of \cite{EFGKK}. At low
temperatures the grand canonical specific heat $c_V^{(g)}=-T\left(\frac{\6^2
  \phi}{\6 T^2}\right)_{\mu,H}$ and the canonical specific
  heat $c_V^{(c)}$ are very similar but at high temperatures differences
appear.

\begin{table}
\begin{ruledtabular}
\begin{tabular}{r| r r r r}
\multicolumn{5} {c}{Specific heat coefficients $U=4$}\\
\hline
$\gamma$ &H=0 & H=0.1& H=0.2 & H=0.3\\
\hline
     n=0.1 & 14.60 [IV] & 1.694 [II] & 1.694 [II] & 1.694  [II]  \\
     n=0.4 & 2.183 [IV] & 2.453 [IV] & 4.117 [IV] & 0.550  [II]  \\
     n=0.7 & 1.396 [IV] & 1.422 [IV] & 1.469 [IV] & 1.556  [IV] \\
     n=0.9 & 1.421 [IV] & 1.440 [IV] & 1.471 [IV] & 1.525  [IV]  \\
\end{tabular}
\end{ruledtabular}
\caption{Specific heat coefficients for $U=4$. The square parentheses identify
the phase of the system at $T=0$. }
\label{tablesh1}
\end{table}
\begin{table}
\begin{ruledtabular}
\begin{tabular}{r| r r r r}
\multicolumn{5} {c}{Specific heat coefficients $U=8$}\\
\hline
$\gamma$ &H=0 & H=0.1& H=0.2 & H=0.3\\
\hline
     n=0.1 & 24.70 [IV] & 1.694 [II] & 1.694 [II] & 1.694  [II]  \\
     n=0.4 & 2.915 [IV] & 4.363 [IV] & 0.550 [II] & 0.550  [II]  \\
     n=0.7 & 1.869 [IV] & 1.954 [IV] & 2.171 [IV] & 2.818  [IV] \\
     n=0.9 & 2.248 [IV] & 2.316 [IV] & 2.469 [IV] & 2.817  [IV]  \\
\end{tabular}
\end{ruledtabular}
\caption{Specific heat coefficients for $U=8$. The square parentheses identify
the phase of the system at $T=0$. }
\label{tablesh2}
\end{table}

In Fig.~\ref{sheatn} we present the temperature dependence of the specific
heat in zero magnetic field for $n=\{0.1,0.4,0.7,0.9,1\}$ and $U=\{0,4,8,
\infty\}$. For fillings $n\ge 0.7$ and small to moderate values of the
interaction strength, $U\le 4$, the specific heat presents a single maximum
which moves to lower temperatures as $U$ increases. For
larger values of the repulsion this single maximum splits into two maxima. The
origin of the lower temperature maximum is due to the spin excitations while
the higher one is due to the charge excitations (which are gapped at
$n=1$). For densities smaller than $0.5$ the specific heat starts to develop a
shoulder which becomes a low temperature maximum even for $U=0$ and $n\sim
[0,0.2].$ The positions and the relative magnitudes of the two maxima depend
in a complex way on the filling factor and the interaction strength. The case
of infinite repulsion is somewhat special. Here we have only one maximum for
$n$ close to 0.5 and the curves for $n=0.1$ and $n=0.9$, which present two
maxima, coincide.  This is due to the fact that for $U=\infty$ and $H=0$ the
system is equivalent to a system of free spins and the symmetry (\ref{symm})
with finite values for $\mu$ stays in the Mott phase with
  fixed particle density 1. For density $n\in[0,1)$ we have
    $c_V^{(c)}(n)=c_V^{(c)}(1-n)$. A similar relation holds for the
  compressibility but not for the magnetic susceptibility.  At half filling
  $c_V^{(c)}\sim0$ for $T>0$.

The temperature dependence of the specific heat in a magnetic field below half
filling is presented in Fig.~\ref{sheath} for $H=\{0,0.1,0.2,0.3\}$ and
$U=\{4,8\}.$ We have chosen values of the magnetic field which are smaller
than $H_0(U)$ which fully polarizes the system even at $n=1$ ($H_0(4)=0.8284$
and $H_0(8)=0.4271$).  The results for $n=0.1$ are almost similar for both
values of $U$ which is easily explainable by the reduced role of the
interaction at low fillings. At $H=0$ the specific heat presents two maxima
the first one being situated at very low temperatures $T<0.005\, t$ and very
large slopes for the linear behavior. Switching the magnetic field fully
polarizes the system and the specific heat coefficient is the same for all
values of $H$. For small values of the magnetic field the two maxima structure
is still present but it disappears for $H=0.3$ becoming a shoulder. Also
compared with a similar structure without the magnetic field the first
maximum is dominant and moves to higher temperature with
$H$. For $n=0.4$ only the $H=0, U=8$ curve presents two maxima and for $T\le
0.025\, t$ $c_V^{(c)}$ is monotonically increasing with $H$. At larger
fillings $n=\{0.7, 0.9\}$ the largest maximum is obtained for zero magnetic
field but at higher temperatures the specific heat again increases with
$H$. The insets of Fig.~\ref{sheath} present the TLL predictions
$c_V^{(c)}=\gamma T$ in the shaded regions and our numerical data outside of
these regions. For each case the right boundary of the shaded regions
represent the lowest temperature accessible with our numerical scheme. For a
given value of the Coulomb repulsion and magnetic field the area of
applicability of the TLL theory is given by the interval of temperature in
which the specific heat is almost linear. From the insets we see that this
interval is largest close to half filling where we have linearity of the
specific heat up to almost $T=0.05\, t$. This interval shrinks dramatically
for low filling fractions. At $n=0.1\, , U=4$ and zero magnetic field the TLL
theory is valid for $T<0.005\, t$ and the interval shrinks even further as we
increase the Coulomb repulsion.

\subsection{Susceptibility}

\begin{figure}
\includegraphics[width=\linewidth]{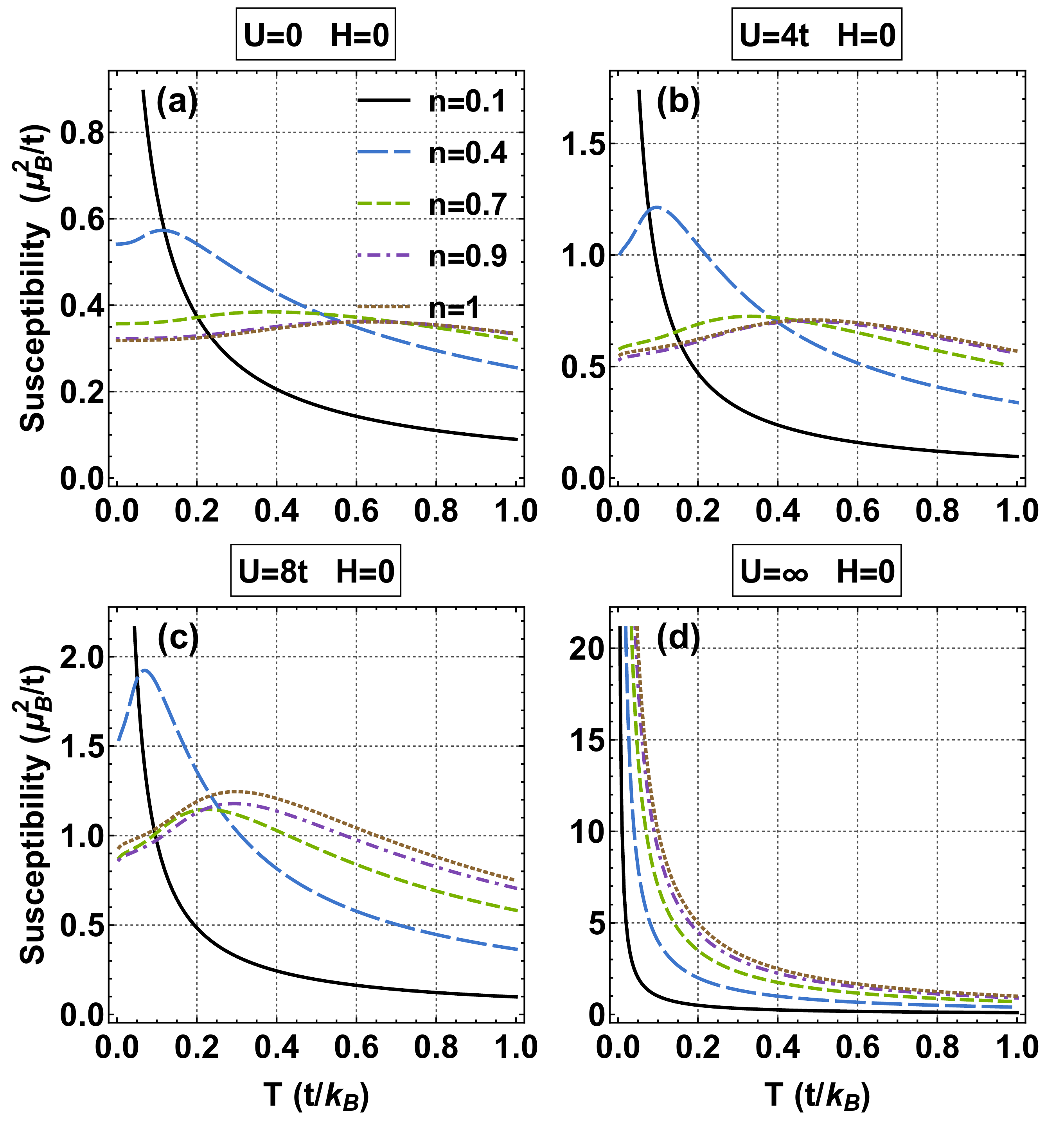}
\caption{Temperature dependence of the susceptibility in zero magnetic field  for various filling
fractions and interaction strengths.
}\label{suscn}
\end{figure}

\begin{figure*}
\includegraphics[width=\linewidth]{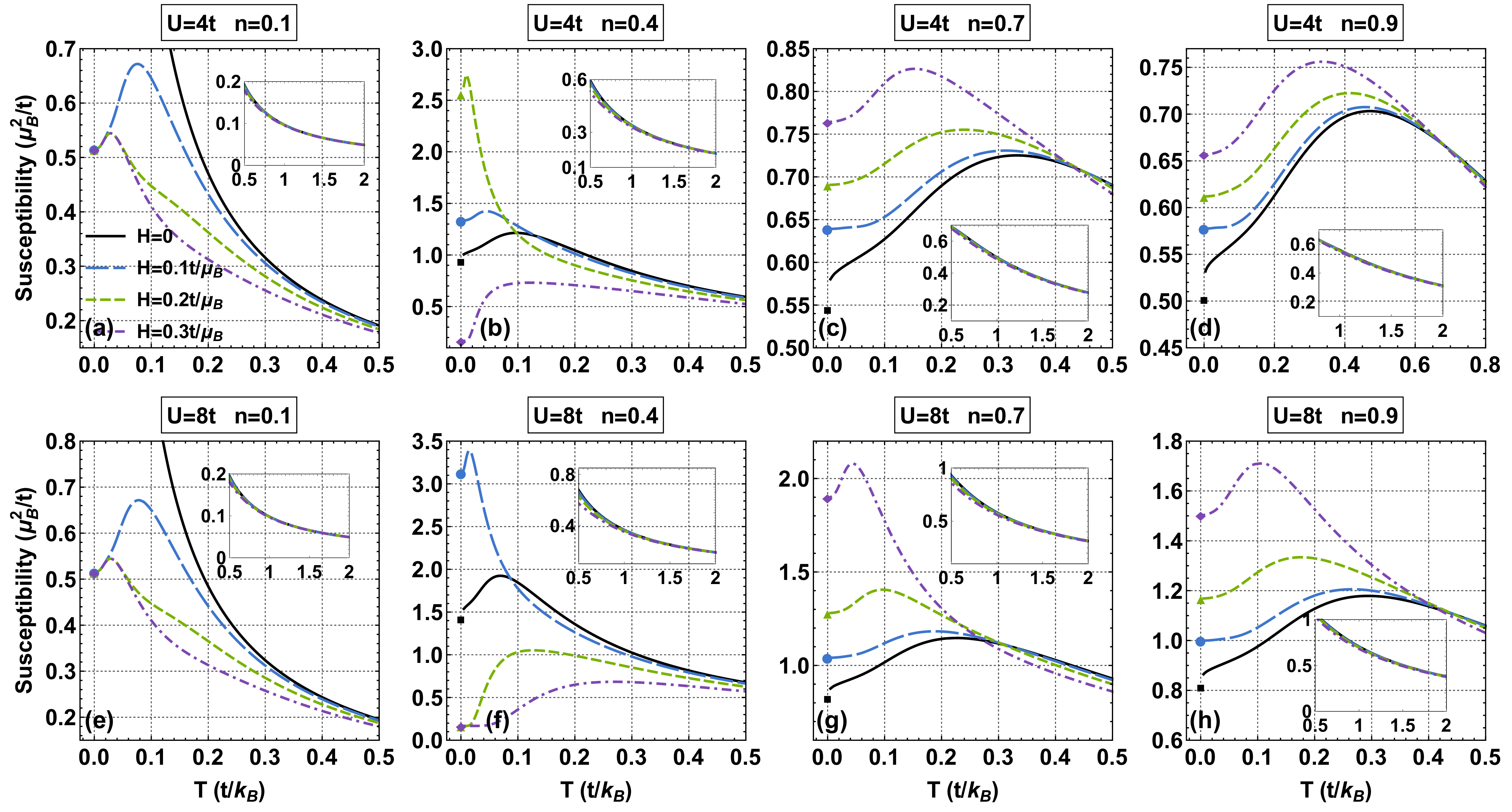}
\caption{Temperature dependence of the susceptibility in the presence of a magnetic field. The values at $T=0$
denoted by squares $(H=0)$, disks $(H=0.1\, t/\mu_B)$, triangles $(H=0.2\, t/\mu_B)$ and diamonds $(H=0.3\, t/\mu_B)$
are computed using Eqs. (\ref{susc2}), (\ref{susc4}) and given in Tables \ref{tablesusc1} and \ref{tablesusc2}.
The insets show the behavior at high temperatures.}
\label{susch}
\end{figure*}
At zero temperature and zero magnetic field the magnetic susceptibility was
investigated in \cite{Tak3,Tak4,Shiba3,UF}. The influence of a magnetic field
was investigated in the very thorough article of Carmelo, Horsch and
Ovchinnikov \cite{CHO} and in \cite{KT1}.  An interesting feature of the
ground state susceptibility is that for all values of the filling fraction and
on-site repulsion it presents a logarithmic dependence at low magnetic fields
$(H\ll 1$) \cite{KT1} i.e.,
\be
\chi(T=0)=\chi_0\left(1+\frac{1}{2\ln (a/H)}-\frac{\ln \ln (a/H)}{4 (\ln (a/H))^2}\right)
\ee
where $a$ is a constant. At finite temperature numerical data for the magnetic susceptibility  can be
found in  \cite{UKO2,KUO1,JKS,Liu,Ha}. In \cite{KT1} (see also \cite{NBTVT}) the authors speculated that the
susceptibility should also have a logarithmic dependence on temperature at any filling fraction and value of $U$,
 a conjecture which is confirmed by our numerical data.

In Fig.~\ref{suscn} we present the temperature dependence of the
susceptibility in zero magnetic field for various filling fractions. For
finite values of the interaction strength the behaviour is similar for all
density values: the susceptibility is finite at $T=0$, presents a maximum
which is inverse proportional with the filling fraction and whose position
moves to higher temperatures as $n$ increases. At higher
temperatures the susceptibility is an increasing function of the filling
fraction. An interesting feature on which we will elaborate later is the
presence of a logarithmic singularity at very low temperatures for
$0<U<\infty$. The case of infinite repulsion is different. Here the
magnetization is $m=n \tanh(H/T)/2$ and the susceptibility is infinite at
$T=0$ for any value of $n$.

\begin{figure}
\includegraphics[width=\linewidth]{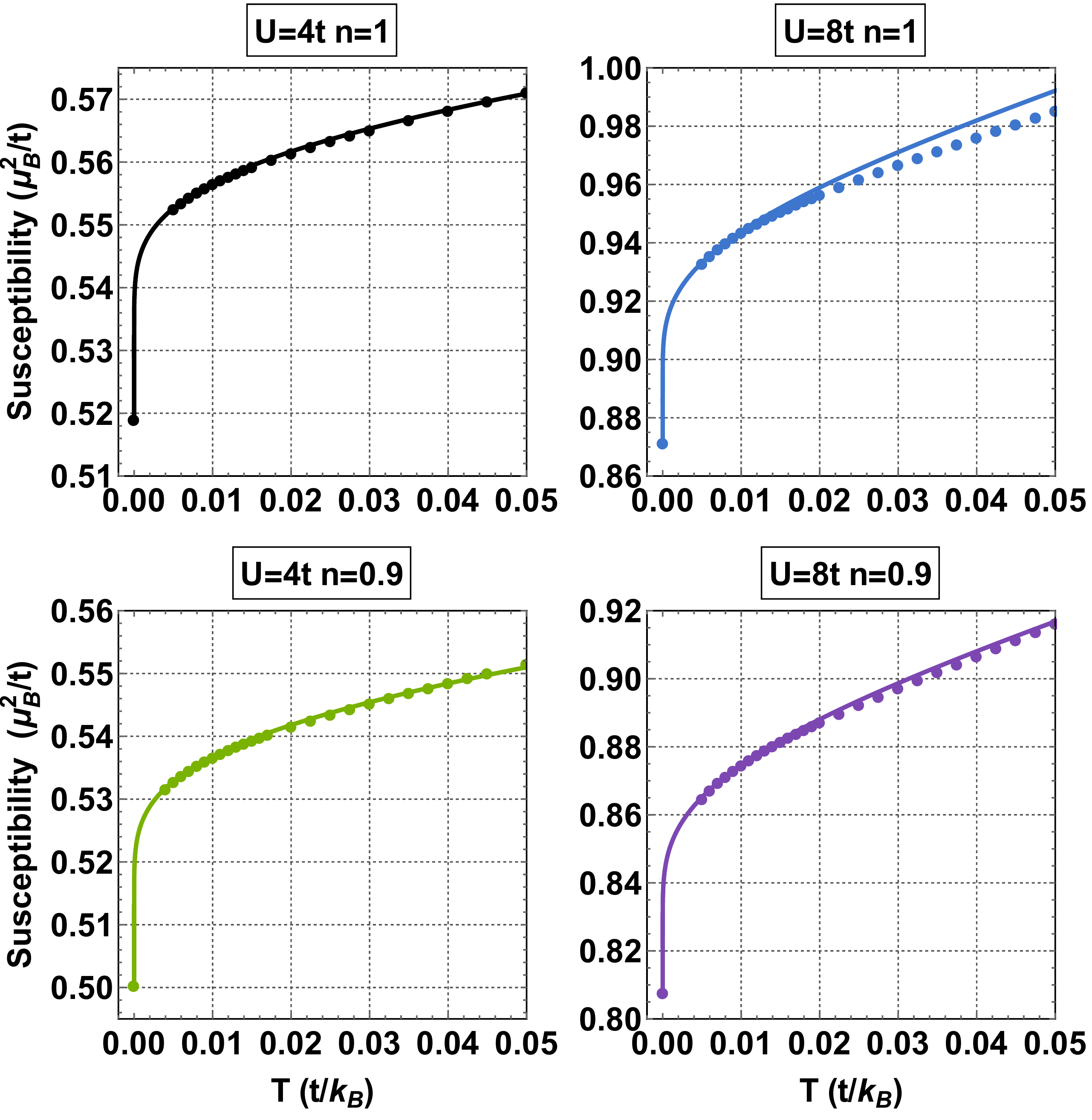}
\caption{Logarithmic dependence of the susceptibility in zero magnetic field. The disks are numerical data
and the lines represent best fits using Eq.~(\ref{logdep}).
}\label{susclog}
\end{figure}

\begin{table}
\begin{ruledtabular}
\begin{tabular}{r| r r r r}
\multicolumn{5} {c}{Susceptibilities at $T=0$ and  $U=4$}\\
\hline
$\chi$ &H=0 & H=0.1& H=0.2 & H=0.3\\
\hline
     n=0.1 & 7.713 [IV] & 0.515 [II] & 0.515 [II] & 0.515  [II]  \\
     n=0.4 & 0.926 [IV] & 1.332 [IV] & 2.561 [IV] & 0.167  [II]  \\
     n=0.7 & 0.543 [IV] & 0.639 [IV] & 0.690 [IV] & 0.764  [IV] \\
     n=0.9 & 0.500 [IV] & 0.577 [IV] & 0.611 [IV] & 0.656  [IV]  \\
\end{tabular}
\end{ruledtabular}
\caption{Susceptibilities at zero temperature and $U=4$. The square
  parentheses identify the phase of the system. Note that we
    are dealing with the grand-canonical spin susceptibility which may be
    non-zero in the spin-polarized phase and independent of the field. In
    fact, this spin susceptibility $\chi$ is identical to the compressibility
    $\kappa$ see e.g. table \ref{tablecomp1}.}
\label{tablesusc1}
\end{table}

\begin{table}
\begin{ruledtabular}
\begin{tabular}{r| r r r r}
\multicolumn{5} {c}{Susceptibilities at $T=0$ and $U=8$}\\
\hline
$\chi$ &H=0 & H=0.1& H=0.2 & H=0.3\\
\hline
     n=0.1 & 13.92 [IV] & 0.515 [II] & 0.515 [II] & 0.515  [II]  \\
     n=0.4 & 1.403 [IV] & 3.123 [IV] & 0.167 [II] & 0.167  [II]  \\
     n=0.7 & 0.816 [IV] & 1.041 [IV] & 1.280 [IV] & 1.897  [IV] \\
     n=0.9 & 0.807 [IV] & 0.999 [IV] & 1.168 [IV] & 1.502  [IV]  \\
\end{tabular}
\end{ruledtabular}
\caption{Susceptibilities at zero temperature and $U=8$. The square parentheses identify
the phase of the system. }
\label{tablesusc2}
\end{table}

The influence of a magnetic field on the temperature dependence of the
susceptibility is presented in Fig.~\ref{susch}.  The general structure is the
same as in the case of zero magnetic field: a finite value at $T=0$ followed
by a maximum at low temperatures. As long as the magnetic field is not strong
enough to fully polarize the system in the ground state the susceptibility
increases with $H$ at low temperatures and the position of the maximum moves
to higher temperatures as the magnetic field decreases (see
panels (c), (d), (g) and (h) of Fig.~\ref{susch}). When the magnetic field
polarizes the system the susceptibility is strongly suppressed and becomes a
monotonically decreasing function of $H$ at low temperatures (see panels (a),
(b), (e) and (f) of Fig.~\ref{susch}).

In both cases at higher temperatures the susceptibility is largest at $H=0$. The most interesting feature is the logarithmic
singularity at low temperatures in zero magnetic field as it can be seen in Fig.~\ref{susclog} where we present our numerical
results together with the  fits using
\be\label{logdep}
\chi(T)=\chi(T=0)+ \frac{a}{\ln(T_0/T)}\, .
\ee
where $a$ and $T_0$ are free parameters. At quarter filling an infinite slope of the susceptibility was obtained from
renormalization group calculations \cite{NBTVT}. The same logarithmic dependence was discovered in the case of the susceptibility
of the XXX spin chain in \cite{EAT,K3}.  While one can say that this phenomenon can be intuitively inferred at half filling
and large  values of $U$ due to the equivalence of the spin sector with the XXX spin chain the logarithmic dependence at
below half filling  which can be seen in the numerical data even at $n=0.1$ and $U=4$ is intriguing.

\subsection{Compressibility}

\begin{figure}
\includegraphics[width=\linewidth]{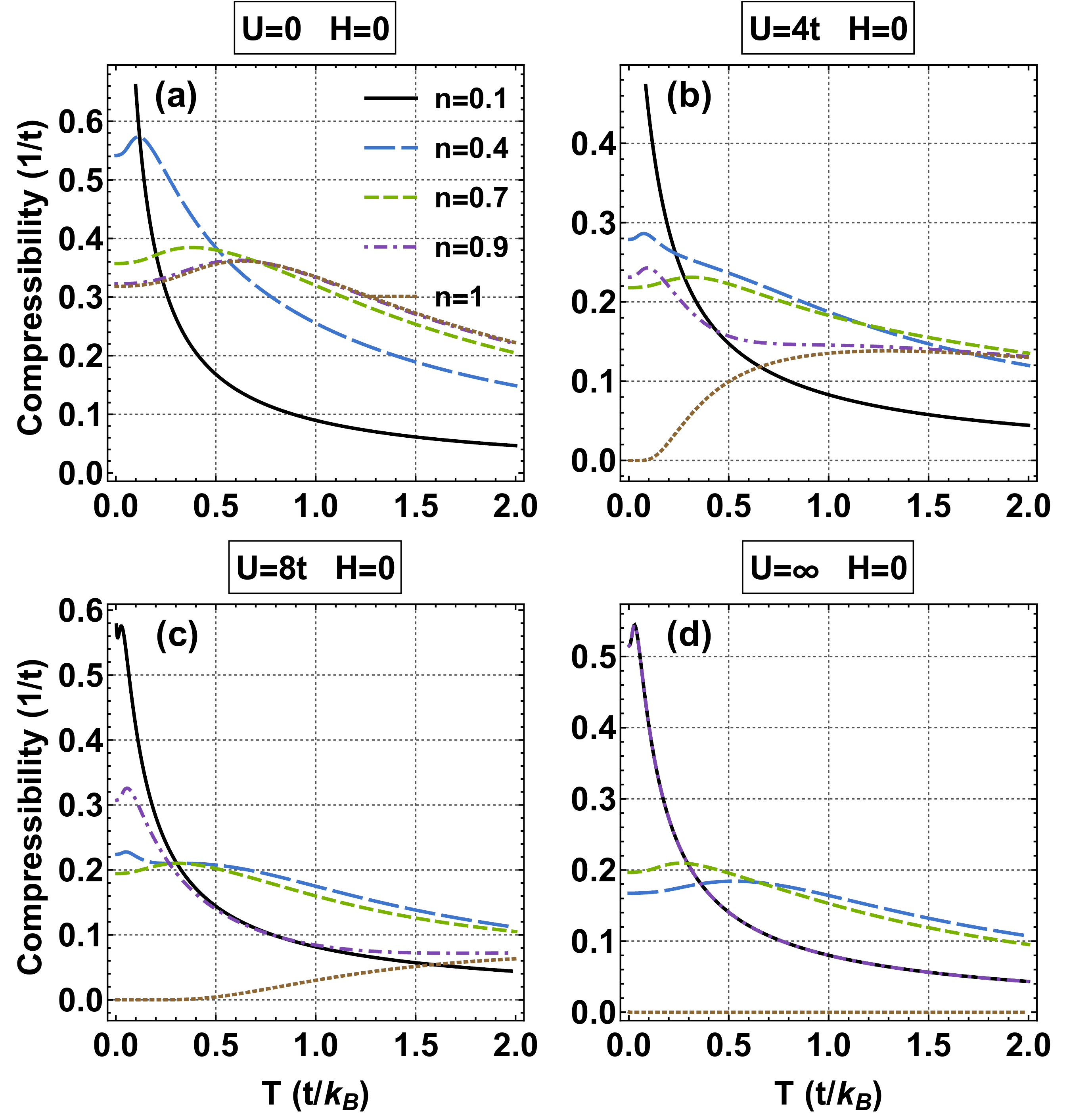}
\caption{Temperature dependence of the compressibility in zero magnetic field  for various filling
fractions and interaction strengths.
}\label{compn}
\end{figure}

\begin{figure*}
\includegraphics[width=\linewidth]{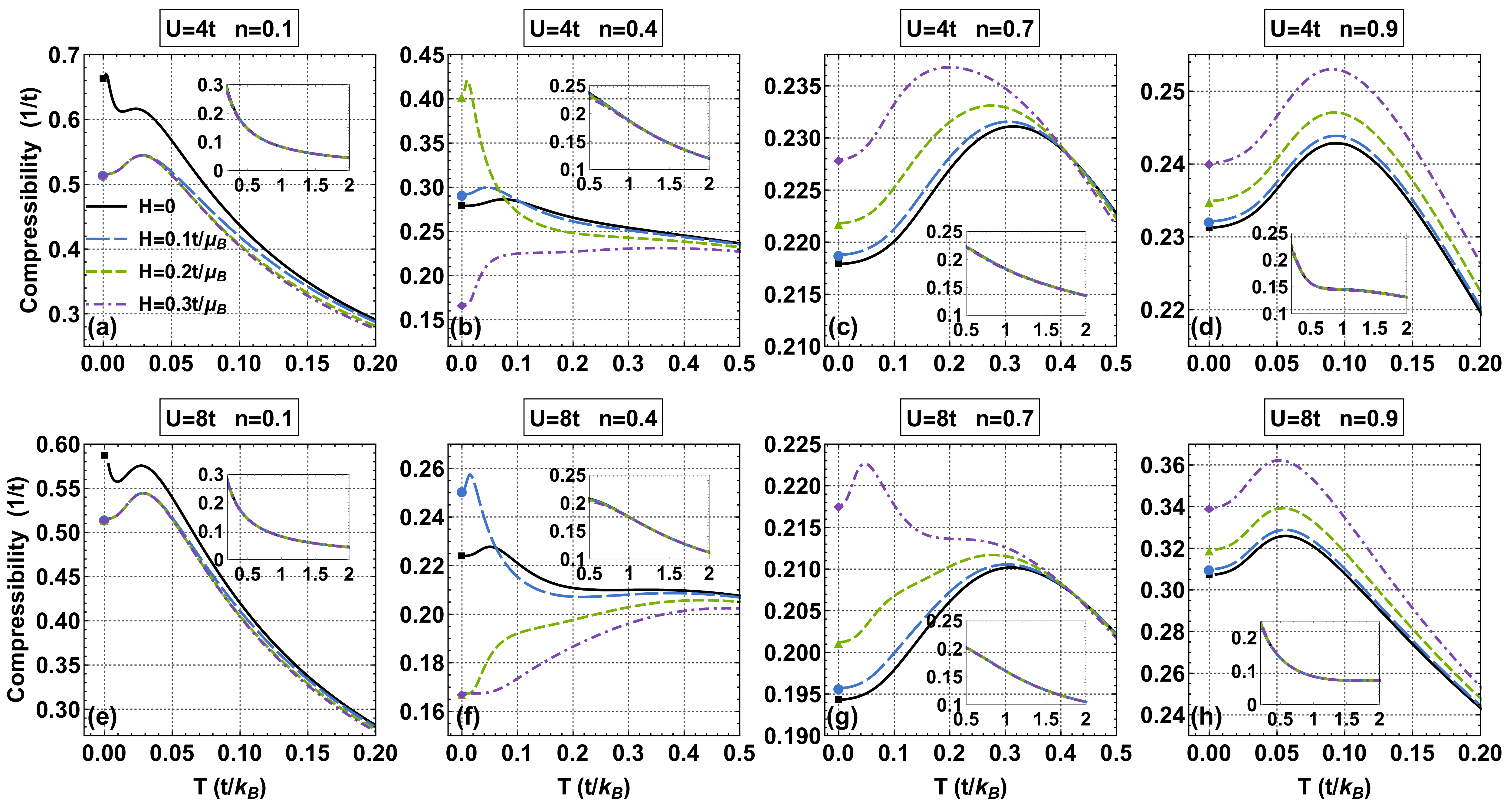}
\caption{Temperature dependence of the compressibility in the presence of a magnetic field. The values at $T=0$
denoted by squares $(H=0)$, disks $(H=0.1\, t/\mu_B)$, triangles $(H=0.2\, t/\mu_B)$ and diamonds $(H=0.3\, t/\mu_B)$
are computed using Eqs. (\ref{susc2}), (\ref{comp4}) and given in Tables \ref{tablecomp1} and \ref{tablecomp2}.
The insets show the behavior at high temperatures.}
\label{comph}
\end{figure*}

The compressibility was investigated analytically and numerically at zero
temperature in \cite{CHO} and at finite temperature in
\cite{UKO2,KUO1,JKS,Ha}.  The temperature dependence of the compressibility in
zero magnetic field and various densities and interaction strengths can be
found in Fig.~\ref{compn}. For $0<U<\infty$ and $n=1$ (half filling) the
charge sector has a gap resulting in zero compressibility at $T=0$ and
exponentially activated behavior at low temperatures.  At arbitrary filling
fractions $\kappa$ is nonzero and in general presents only one
maximum. However at very low fillings the compressibility is strongly enhanced
and it can present two maxima at very low temperatures.  Similar to the case
of the specific heat at $U=\infty$ the compressibility satisfies
$\kappa(n)=\kappa(1-n)$ for $n\in[0,1)$ (the curves for $n=0.1$ and $n=0.9$
  coincide as in Fig.~\ref{sheatn}).

The influence of the magnetic field on compressibility is shown in Fig.~\ref{comph}. The presence of a magnetic field
which polarizes the system at $n=0.1$ depresses the compressibility below the zero field value. In addition $\kappa$
presents only one maximum. Magnetic fields below the critical value in general enhance the compressibility and move
the maximum to lower temperatures. For densities close to $n\sim 0.5$ large values of $H$ seem to develop a second
maximum at low temperatures similar to the case of low fillings but a similar phenomenon was reported also for
zero magnetic field in \cite{JKS}.

\begin{table}
\begin{ruledtabular}
\begin{tabular}{r| r r r r}
\multicolumn{5} {c}{Compressibilities at $T=0$ and  $U=4$}\\
\hline
$\kappa$ &H=0 & H=0.1& H=0.2 & H=0.3\\
\hline
     n=0.1 & 0.662 [IV] & 0.515 [II] & 0.515 [II] & 0.515  [II]  \\
     n=0.4 & 0.278 [IV] & 0.291 [IV] & 0.403 [IV] & 0.167  [II]  \\
     n=0.7 & 0.217 [IV] & 0.218 [IV] & 0.221 [IV] & 0.227  [IV] \\
     n=0.9 & 0.231 [IV] & 0.232 [IV] & 0.234 [IV] & 0.240  [IV]  \\
\end{tabular}
\end{ruledtabular}
\caption{Compressibilities at zero temperature and $U=4$. The square parentheses identify
the phase of the system. }
\label{tablecomp1}
\end{table}

\begin{table}
\begin{ruledtabular}
\begin{tabular}{r| r r r r}
\multicolumn{5} {c}{Compressibilities at $T=0$ and $U=8$}\\
\hline
$\kappa$ &H=0 & H=0.1& H=0.2 & H=0.3\\
\hline
     n=0.1 & 0.587 [IV] & 0.515 [II] & 0.515 [II] & 0.515  [II]  \\
     n=0.4 & 0.223 [IV] & 0.250 [IV] & 0.167 [II] & 0.167  [II]  \\
     n=0.7 & 0.194 [IV] & 0.195 [IV] & 0.201 [IV] & 0.217  [IV] \\
     n=0.9 & 0.307 [IV] & 0.309 [IV] & 0.319 [IV] & 0.339  [IV]  \\
\end{tabular}
\end{ruledtabular}
\caption{Compressibilities at zero temperature and $U=8$. The square parentheses identify
the phase of the system. }
\label{tablecomp2}
\end{table}

\section{Thermodynamic properties at half filling}\label{s6}

\begin{figure}
\includegraphics[width=\linewidth]{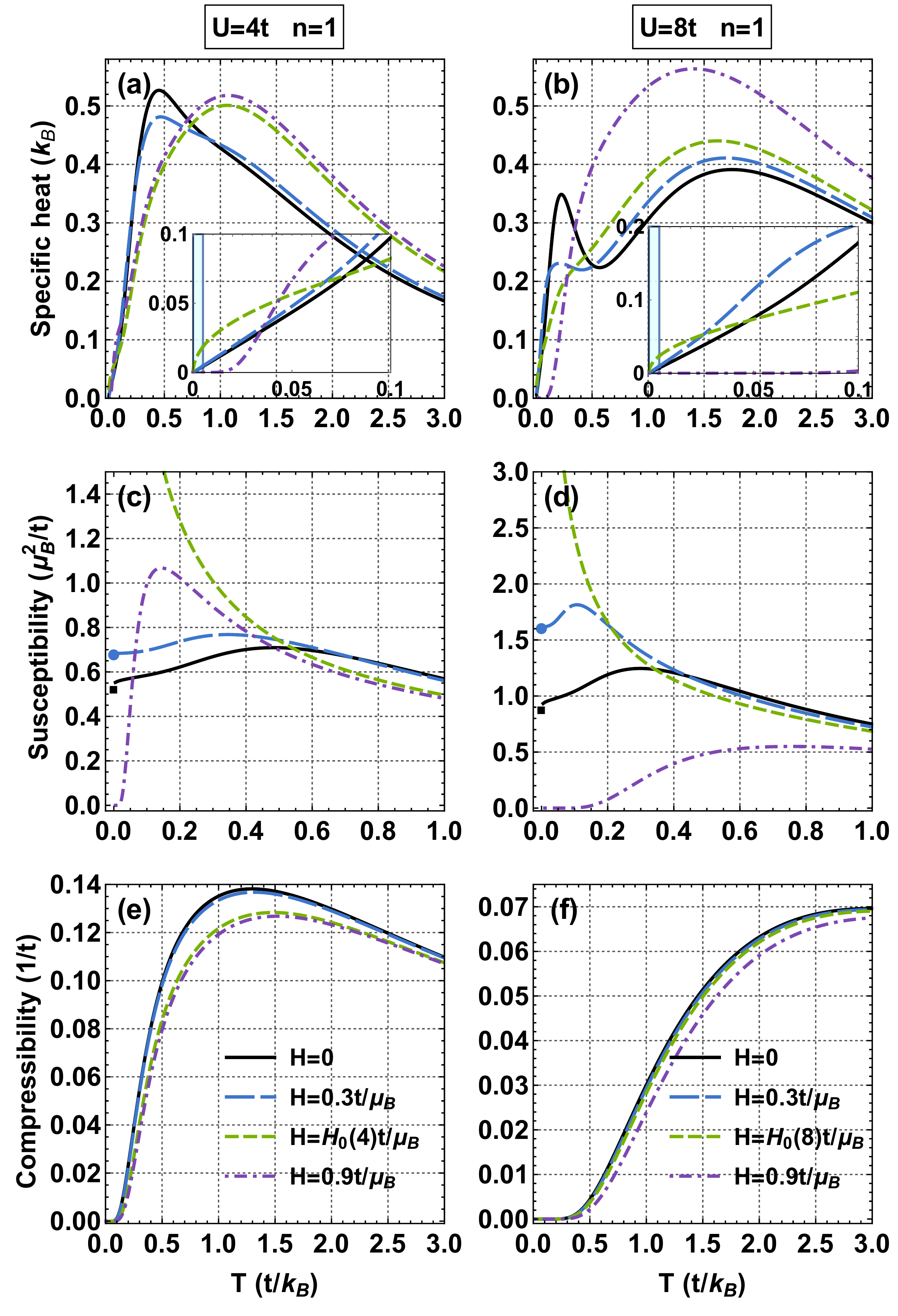}
\caption{Temperature dependence of the specific heat, susceptibility and compressibility at half
filling in the presence of a magnetic field.  The insets present a zoom of the specific heat
data at low temperatures. The shaded regions are inaccessible numerically and contain the theoretical
predictions $c_V^{(c)}\sim\gamma T$ for $H<H_0(U),$ $c_V^{(c)}\sim a T^{1/2}$ for $H=H_0(U)\, t/\mu_B$  and
$c_V^{(c)}\sim T^{3/2} e^{-\alpha/T}$ for $H=0.9\, t/\mu_B.$ The susceptibilities at $T=0$ denoted by squares
$(H=0)$ and disks $(H=0.3\, t/\mu_B)$ are computed using Eq.~(\ref{susc5}).
 }
\label{hf}
\end{figure}

Historically the half filling case constitutes one of the most investigated regimes of the Hubbard model starting
with the initial paper of Lieb and Wu \cite{LW}.  At half filling the charge sector has a gap and therefore the
low temperature thermodynamics is influenced only by the spin excitations. At zero magnetic field and temperature
the dressed charge for the half filled band is $\xi(A=\infty)=1/\sqrt{2}$ \cite{WE1} resulting in a specific heat
coefficient and susceptibility (the compressibility is zero as a consequence of the charge gap)
\be
\gamma_{\mbox{\tiny{HF}}} (H=0)=\frac{\pi}{3}\frac{1}{v_s}\, , \ \ \  \chi_{\mbox{\tiny{HF}}} (H=0)=\frac{4}{2\pi v_s}\, ,
\ee
with the spin velocity
\be
v_s= 2I_1\left(\frac{\pi}{2 u}\right)/I_0\left(\frac{\pi}{2 u}\right)\, ,
\ee
where $I_n(z)$ is the modified Bessel function of the first kind which for integer $n$ has the
integral representation $I_n(z)=\frac{1}{\pi}\int_{0}^\pi e^{z \cos \theta}\cos(n \theta)\, d\theta\, .$
For magnetic fields $H<H_0(U)$ the system is partially polarized (phase V) and for $H>H_0(U)$
is fully polarized (phase III).

In Fig.~\ref{hf} (a) and (b) we present the temperature dependence of the specific heat for $U=4$  and $U=8$
and several values of the magnetic $H=\{0,0.3\}<H_0(U),$  $H=H_0(U)$  and $H=0.9>H_0(U)$. At low and moderate interaction
strengths and magnetic field below $H<H_0(U)$ the specific heat is linear in temperature $(\gamma(U=4,H=0)=0.853\, ,
\gamma(U=4,H=0.3)=0.933)$ and presents a single maximum which decreases with the increase of $H$. The results for
$U=8$ show the double peak structure with the second maximum being proportional to the magnetic field ($(\gamma(U=8,H=0)=1.433\, ,
\gamma(U=8,H=0.3)=1.928)$.
The specific heat coefficients monotonically increase as $H$ approaches $H_0(U)$ (the spin velocity decreases) the value for
which they become infinite signalling the QPT from phase V to phase III. At the critical field the specific heat
behaves like $c_V^{(c)}\sim a(U)\, T^{1/2}$  with $a(U=4)=0.261$ and $a(U=8)=0.332$ and presents only one maximum.
For $H=0.9>H_0(U)$ the system presents thermally activated specific heat $c_V^{(c)}\sim T^{3/2} e^{-\alpha/T}$.

Results for the susceptibility are shown in Fig.~\ref{hf} (c) and (d). For
$H=0$ and $H=0.3$ the curves present only one maximum which moves
to lower temperatures with increasing magnetic field and the
zero field susceptibility presents logarithmic behaviour for low temperatures
(see Fig.~\ref{susclog}).  For $T=0$ and magnetic fields lower than but close
to $H_0(U)$ the susceptibility diverges like $1/(H_0-H)^{1/2}$ and for
$H>H_0(U)$ vanishes. As a function of the temperature $\chi$ is divergent at
low-$T$ and $H=H_0(U)$ and behaves like $e^{-\beta/T}$ for $H>H_0(U)$.

At half filling and zero temperature the compressibility is zero for all values of the magnetic field.
At low temperatures it behaves like $e^{-\beta/T}$  and presents a single maximum which moves to higher
temperatures as $U$ is increased (Fig.~\ref{hf} (e) and (f)). At half filling the compressibility
is a decreasing function of the magnetic field for all temperatures.

\section{Double occupancy}\label{s7}

\begin{figure}
\includegraphics[width=\linewidth]{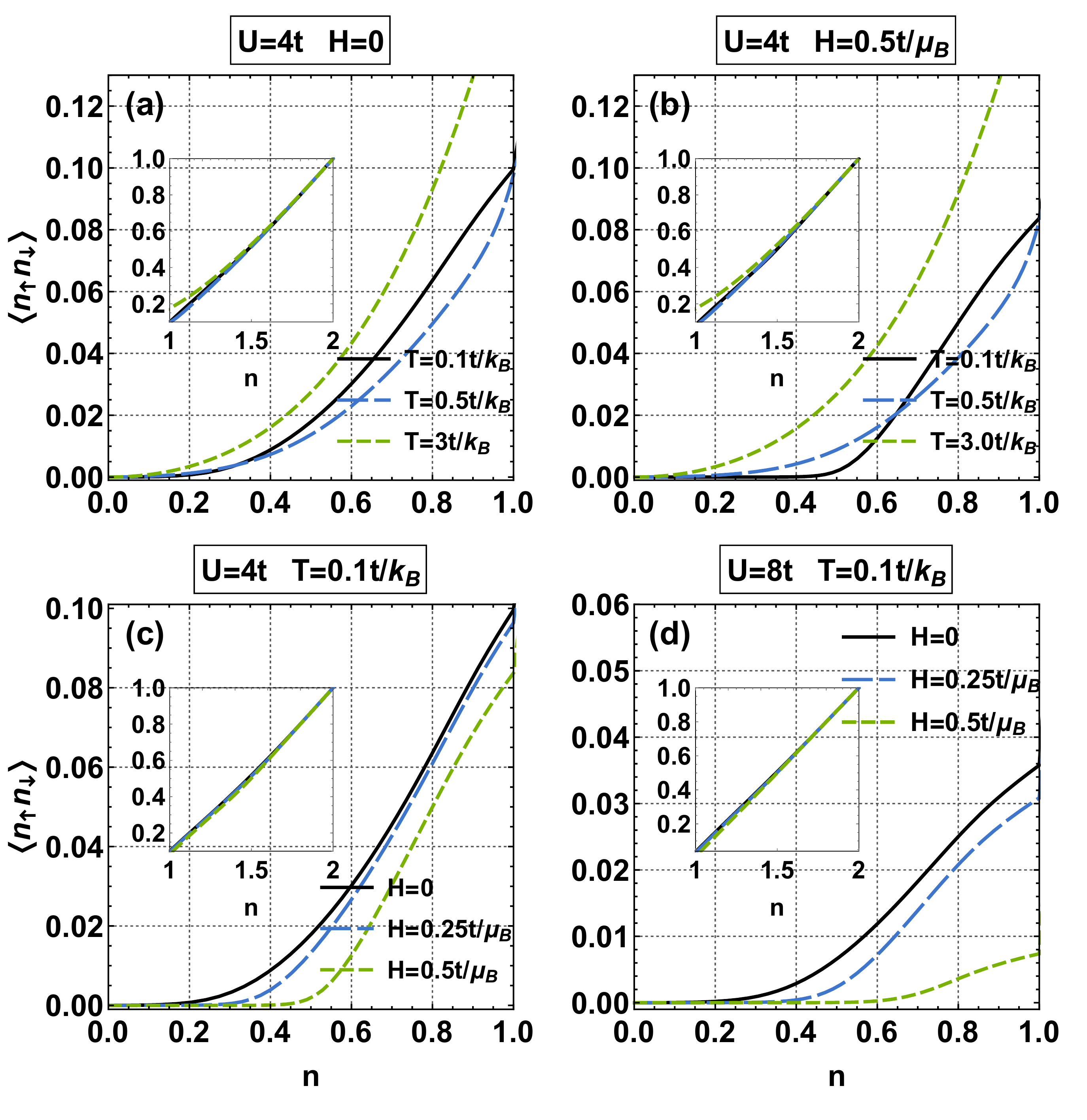}
\caption{Double occupancy as a function of the density for different values of temperature,  magnetic fields
and $U=\{4t,8t\}.$  The insets contain the results for the density in the $(1,2]$ interval.
 }\label{do}
\end{figure}

The computation of the correlation functions in the Hubbard model is an
extremely difficult task even though in principle the wave functions and
energy levels are known. However, one particular correlation function, which
is also experimentally accessible, the double occupancy, can be determined
from the thermodynamics of the system in a manner similar with the derivation
of the contact in the case of continuous short range interaction models
\cite{Tan1,Tan2,Tan3,OD,BP1,ZL,WC1,WC2,VZM1,VZM2, BZ,PK2}.  The double
occupancy at site $j$ defined by $\langle n_{j,\uparrow}
n_{j,\downarrow}\rangle$ quantifies the probability that a lattice site has
two electrons.  Due to translation invariance we have $\langle n_{j,\uparrow}
n_{j,\downarrow}\rangle=\langle n_{k,\uparrow} n_{k,\downarrow}\rangle$ for
any $j,k\in\{1,\cdots,L\}$ and we define $\langle n_{\uparrow}
n_{\downarrow}\rangle =\sum_{j=1}^L \langle n_{j,\uparrow}
n_{j,\downarrow}\rangle/L$. Using the definition of the grand canonical
potential $\phi=-\ln Z/(\beta L)$ with $Z= \mbox{Tr}\left[ e^{-\beta
    \mathcal{H}-\mu N - 2H m}\right]$ and the Helmann-Feynman theorem we find
$\left(\frac{\6 \phi}{\6 U}\right)_{\mu,H,T}= \mbox{Tr} \left[\sum_{j=1}^L
  \left(n_{j,\uparrow}-\scalebox{0.8}{$\frac{1}{2}$}\right)\left(n_{j,\downarrow}-\scalebox{0.8}{$\frac{1}{2}$}\right)
  e^{-\beta \mathcal{H}-\mu N - 2H m}\right]/(LZ)$ and therefore
\be\label{defdo}
\langle n_{\uparrow} n_{\downarrow}\rangle=\left(\frac{\6 \phi}{\6 U}\right)_{\mu,H,T}+\frac{n}{2}-\frac{1}{4}\, .
\ee
At half filling the double occupancy was investigated  in  \cite{Shiba1,Shiba2,GKS,GRPSK,CCHQS,STUBG} and for the
spin-disordered regime in \cite{EEG}. The dependence on the temperature at any filling and zero magnetic field
was studied by Campo \cite{Campo} using the QTM equations.

\begin{figure}
\includegraphics[width=\linewidth]{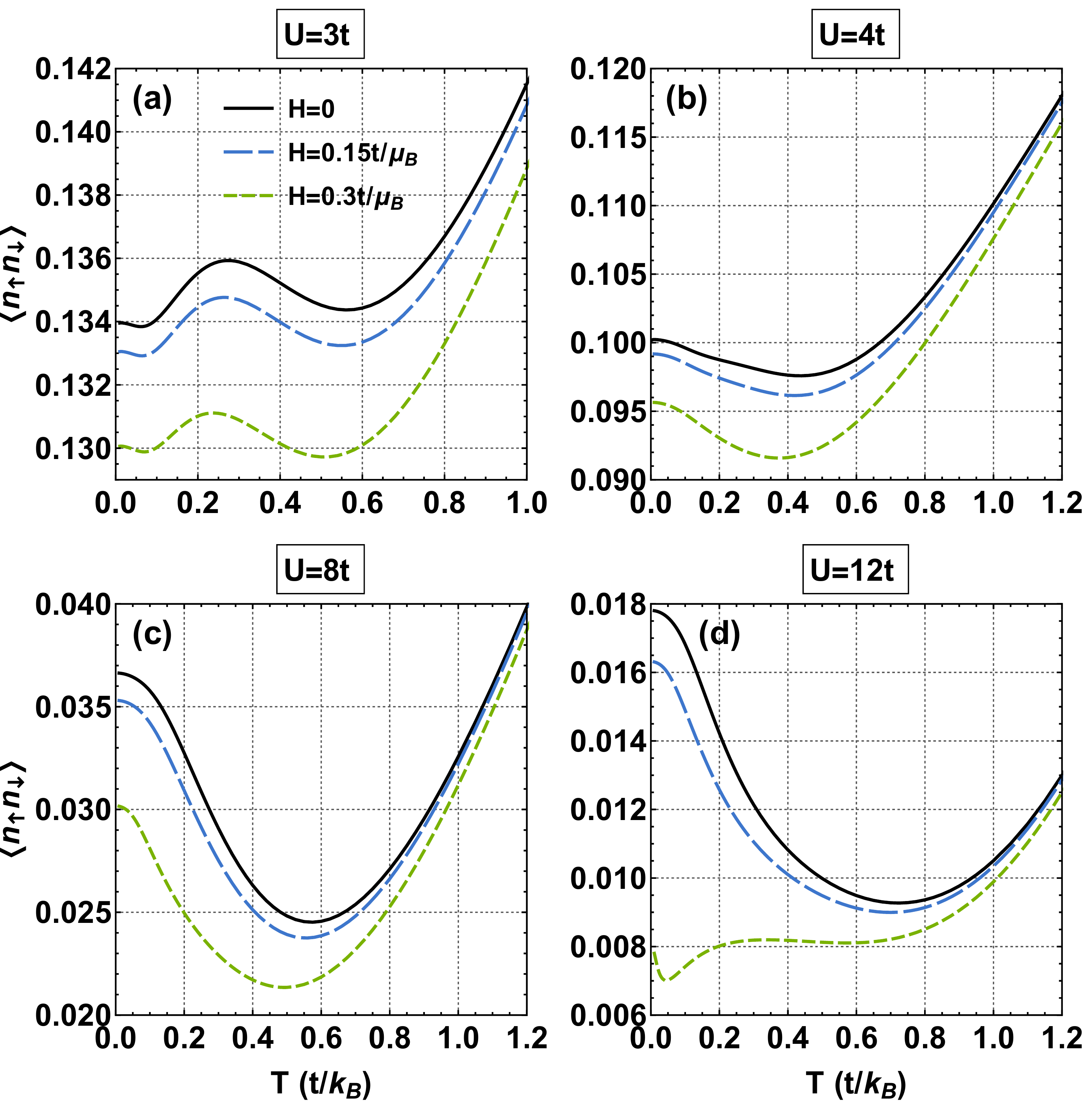}
\caption{Double occupancy at half filling as a function of the temperature for different values of $U$ and magnetic fields
$H=\{0, 0.15\, t/\mu_B, 0.3\, t/\mu_B\}.$  The double occupancy presents two minima for intermediate values of $U$ (a)  and
for large values of $U$  and magnetic field close to the critical value (d) ($H_0(12)\sim 0.3245 \, t/\mu_B$).
 }\label{dop}
\end{figure}

The double occupancy $d(n)$ takes values between 0 and 1 and unlike many other
thermodynamic functions it is not symmetric with respect to $n=1$.
From (\ref{symm}) we find by differentiation with respect to $u$ the
relation $d(n)=d(2-n)+n-1$.
At zero temperature it is a continuous function of $n\in[0,2]$
  but the derivative is discontinuous at half filling. Numerical results for
  different temperatures and magnetic fields are presented in Fig.~\ref{do}.
  For free particles on the lattice one would expect that $\langle
  n_{\uparrow} n_{\downarrow}\rangle$ would monotonically increase with
  temperature. In the Hubbard model due to the repulsive interaction term this
  simple picture does not hold. At zero magnetic field and $U=4$ (panel (a) of
  Fig.~\ref{do}) the free particle picture holds for small filling fractions
  where the interaction is not that important. For $n\sim 0.3$ a crossover
  appears and the double occupancy for $T=0.1$ is larger than the equivalent
  quantity at $T=0.5$ for $n\in(0.3,1)$. Switching a magnetic field (panel (b)
  of Fig.~\ref{do}) moves the crossover density closer to half filling. At
  fixed temperature the magnetic field suppresses the double occupancy as it
  can be seen in Fig.~\ref{do} (c) and (d). This is due to the fact that the
  magnetic field polarizes the system and therefore the probability to have
  two electrons is doubly penalized by the Pauli principle and the repulsive
  interaction.  Also at fixed temperature $\langle n_{\uparrow}
  n_{\downarrow}\rangle$ is a monotonically decreasing function of interaction
  strength and magnetic field as the system becomes more easily polarizable as
  $U$ increases.

At half filling and zero temperature and magnetic field the double occupancy can be obtained in analytic form from
the ground state energy of Lieb and Wu
\be
\langle n_{\uparrow} n_{\downarrow}\rangle_{0}(U)=\frac{1}{2}\int_{0}^\infty d\omega J_0(\omega)J_1(\omega) \mbox{sech}^2 (\omega U/4)\, ,
\ee
with $J_n(\omega)=\frac{1}{\pi}\int_{0}^\pi \cos(\omega \sin \theta-n\theta)\, d\theta$ the $n$th Bessel function of the first kind.
At low temperatures $(T\ll t)$ we have \cite{CCHQS, KBar}
\be
\langle n_{\uparrow} n_{\downarrow}\rangle(T) \underset{T\ll t}{\simeq} \langle n_{\uparrow} n_{\downarrow}\rangle_{0}(U)-\frac{1}{2}\mathcal{C}(2\pi/U) T^2 +O(T^3)
\ee
with
\be\label{funcc}
\mathcal{C}(x)=\frac{x^2}{12}\left(1-\frac{I_0(x)[I_0(x)+I_2(x)]}{I_1(x)^2}\right)
\ee
and at high temperatures $(T> t^2/U)$ the following expansion is valid \cite{CCHQS,TS}
\begin{align}
\langle n_{\uparrow} n_{\downarrow}\rangle(T)& \underset{T> t^2/U}{\simeq}
-\frac{1}{2}\left(-T U +T U \cosh(U/2T)\right)^{-1}\nonumber\\
&+\tanh(U/4T)\mbox{sech}^2(U/4T)/(8T^2)\nonumber\\
&+\frac{1}{2}\frac{1-4U^{-2}}{e^{U/2T}+1}+1/U^2+O(T^{-4})\, .
\end{align}

At half filling the dependence of the double occupancy on the temperature in
the presence of a magnetic field for $U=\{3,4,8,12\}$ is shown in
Fig.~\ref{dop}. At low temperatures and for all values of the coupling
strength the double occupancy decreases with increasing temperature. This is
relatively counterintuitive but it facilitates spin excitations which are
entropically preferred (also notice that at zero magnetic field the function
$\mathcal{C}(x)$ is positive for all $U$ \cite{CCHQS}). It is easy to see from
(\ref{defdo}) and the definition of the entropy that at half filling these
quantities satisfy the Maxwell relation $\6 s/\6 U=-\6 \langle n_{\uparrow}
n_{\downarrow}\rangle/\6 T$ which shows that the decrease of the double
occupancy is accompanied by an increase of the entropy which is an analog of
the Pomeranchuk effect \cite{Pomer,Rich}. For large values of the on-site
repulsion ($U\ge 4$) and moderate magnetic fields the double occupancy
presents only one minimum which moves to lower temperatures
with increasing $H$.  At intermediate values of $U$ it can be seen in
Fig.~(\ref{dop})(a) that $\langle n_{\uparrow} n_{\downarrow}\rangle$ presents
a two minima phenomenon which was first reported in \cite{STUBG} in the case
of zero magnetic field. Here we show that a similar doubly nonmonotonic (two local minima)
behavior is present at large interaction strengths and magnetic fields close
to the critical value $H_0(U)$ (see Fig.~\ref{dop} (d)). Also in this case the
specific heat presents three maxima instead of two as in the case of zero or
moderate magnetic field. For all values of the interaction strength and half
filling the double occupancy is a monotonically decreasing function of the
magnetic field.

\begin{figure*}
\includegraphics[width=\linewidth]{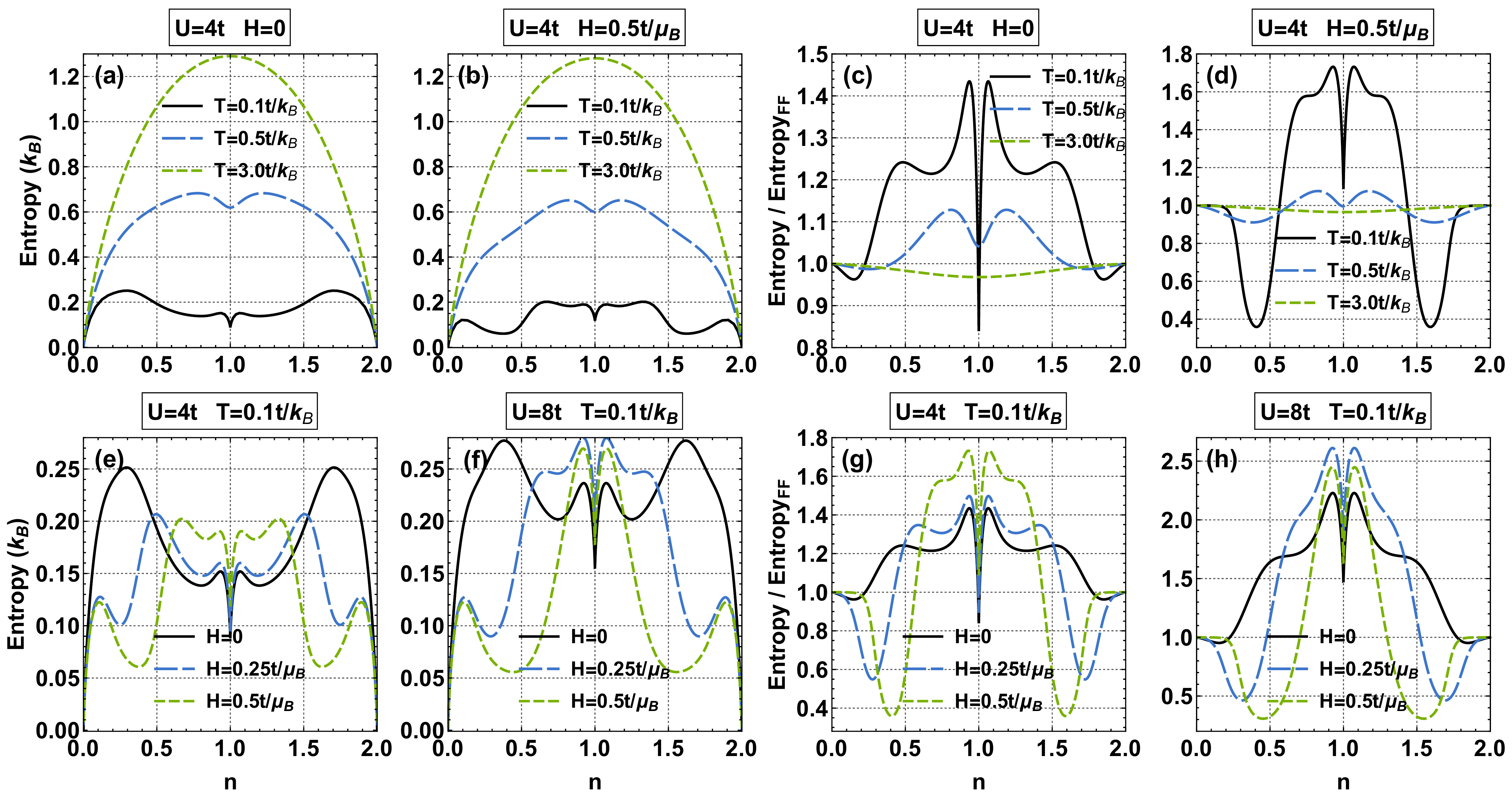}
\caption{(a), (b), (e), (f): Dependence of the entropy on the density for  different temperatures and magnetic fields.
(c), (d), (g), (h): Dependence of the ratio of the entropy of the Hubbard model to the entropy of lattice free fermions
on the filling factor.
}
\label{entropy}
\end{figure*}

\section{Entropy}\label{s8}

In the left panels of Fig.~\ref{entropy} we present the dependence of the entropy $s=-\left(\frac{\6 \phi}{\6 T}\right)_{\mu,H}$ on
the density for different temperatures and magnetic fields. Similar results at zero magnetic field were reported in \cite{Campo}.
The right panels contain the ratios between the entropy of the Hubbard
model and that of a system of free fermions at the same density. The dependence of the entropy on the filling factor is highly
nonmonotonic at low temperatures. For $U=4$ and zero magnetic field  the $T=0.1$ curve presents two local maxima in the $[0,1]$ interval
and  minima at endpoints (the entropy is symmetric in $n$ with respect to $n=1$). Increasing the temperature the dependence
becomes smoother with the well known high temperature enveloping curve
\be
s(T\gg1)=2\ln\left(\frac{2}{2-n}\right)-n \ln\left(\frac{n}{2-n}\right)\, .
\ee
The origin of the two maxima is due to the quantum phase transitions, phase I to phase IV, for the
first maximum and phase IV to phase V for the second one. In the presence of a magnetic field the
entropy presents three maxima at low temperatures as a consequence of the three QPTs (phase I to phase II,
phase II to phase IV and phase IV to phase V) present in the phase diagram for $H<H_0(U)$.
For a fixed temperature the magnetic field decreases the entropy at low fillings compared with the $H=0$
case and close to half filling the effect is inverse (see Fig.~\ref{entropy} panels (e) and (f)). It should
be noted that for $U=8$, $H=0.5$ is larger than $H_0(8)$ and therefore the entropy presents only two maxima
as the system has only two QPTs (phase I to phase II and phase II to phase III).

The ratio $s/s_{FF}$ with $s_{FF}$ the entropy of a system of lattice free fermions at the same density  measures
the deviations from the free system. We see in Fig.~\ref{entropy} (c), (d), (g) and (h) that this ratio presents large
deviations from 1 as we increase the interaction strength (as expected) but also as we increase the magnetic field
as long as $H<H_0(U)$.

\section{Density profiles}\label{s9}

\begin{figure}
\includegraphics[width=\linewidth]{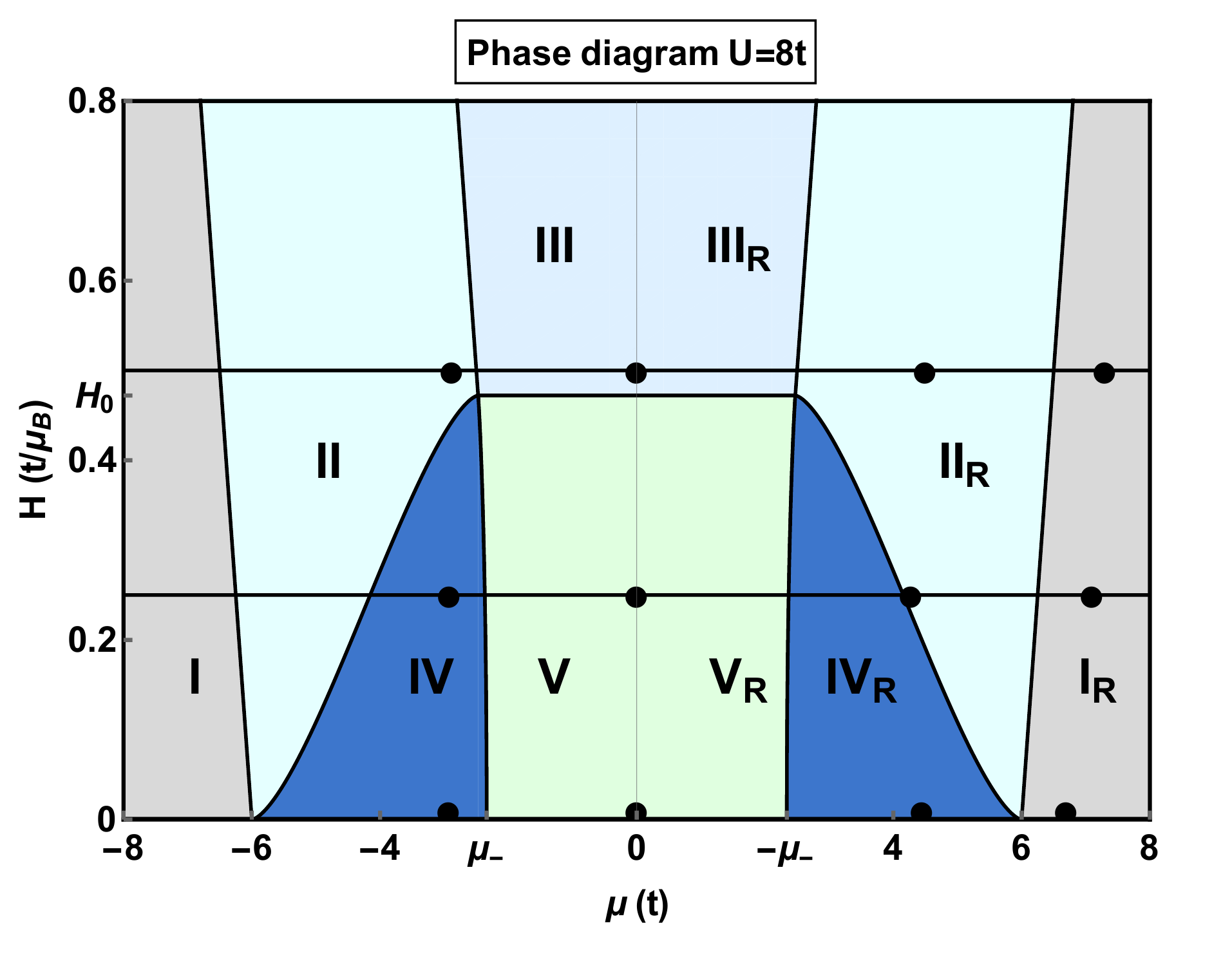}
\caption{Phase diagram at zero temperature for $U=8$ and both positive and negative values of the chemical potential.
The three horizontal  lines pass through $H=\{0, 0.25\, t/\mu_B, 0.5\, t/\mu_B\}$ and the disks on each one correspond to  values of the
chemical potential  for which the density is (from left to right) $n=\{0.8, 1, 1.5, 2\}$.
}\label{pddp}
\end{figure}
\begin{figure*}
\includegraphics[width=\linewidth]{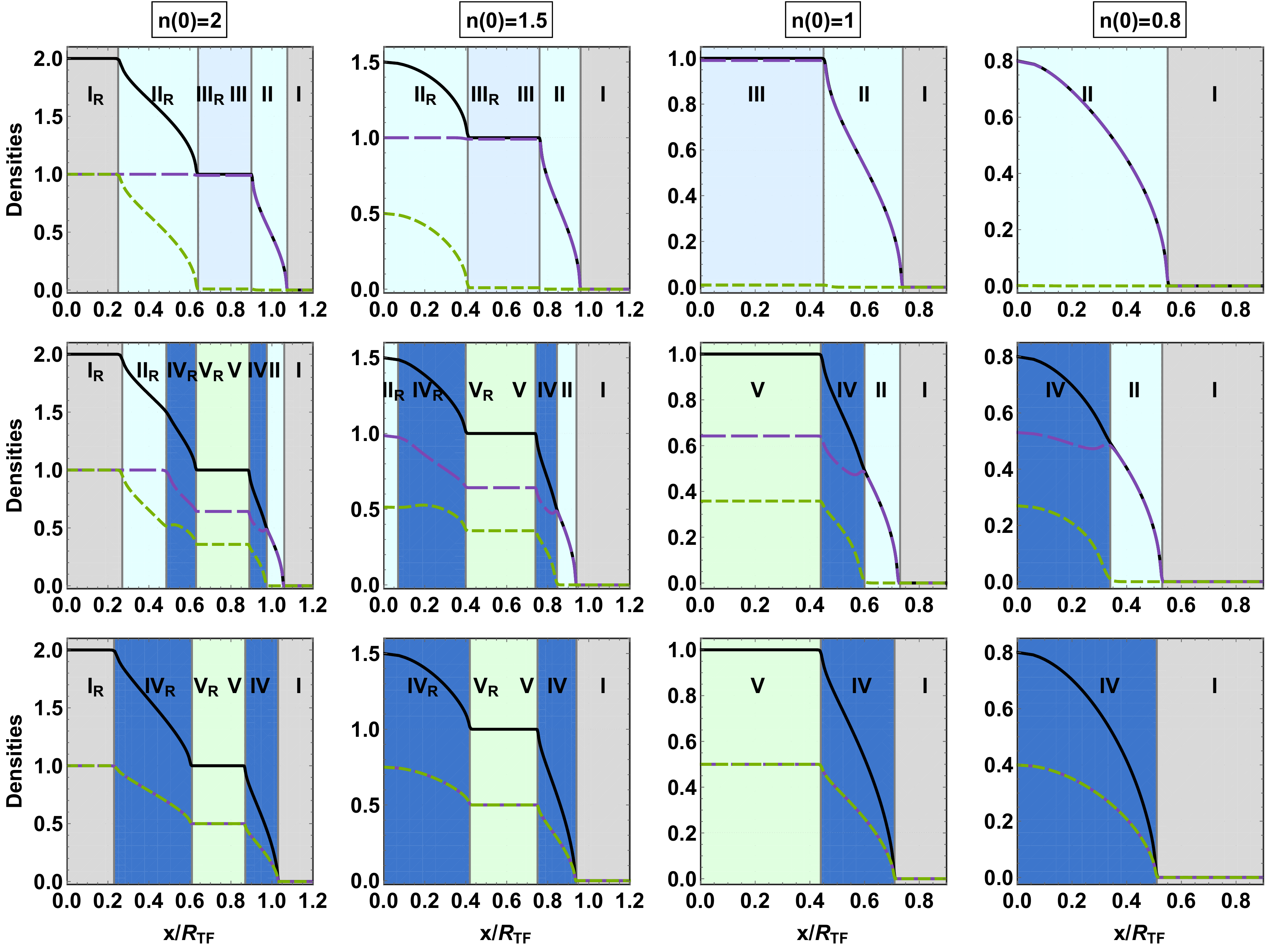}
\caption{Density profiles for the Hubbard model in the presence of a trapping potential at $T=0.025 t$, $U=8 t$ and $H=0.5\, t/\mu_B$
(top row), $H=0.25 \, t/\mu_B$ (middle row) and $H=0$ (bottom row). The profiles are symmetric with respect to $x$. Here we present
only the $x>0$ region. At this temperature the density profiles are effectively  indistinguishable from the ones at $T=0$.
The black, violet and green lines represent the total density  $n(x)=n_\uparrow(x)+n_\downarrow(x)$,  the density of spin up
electrons $n_\uparrow(x)$  and the density of spin down electrons  $n_\downarrow(x)$. Each phase of the system is highlighted
in the same color as in the phase diagram presented in Fig.~\ref{pddp}.
}
\label{densp}
\end{figure*}

The experimental realization of various physical models of interest using ultracold atoms in optical lattices
require a trapping potential. In the case of the Hubbard model the presence of a harmonic trap is equivalent
with the addition to the Hamiltonian (\ref{ham}) of a site dependent potential of the type
\be
\mathcal{H}_{trap}=\sum_{j,\sigma} \frac{m \omega_{e}^2}{2} j^2 n_{j\sigma}\, ,
\ee
with $\omega_{e}$ the effective trap frequency. The addition of such a term breaks the integrability of the Hubbard
model but considering a large system and a slowly varying trapping potential the thermodynamic properties
of the inhomogeneous system can be computed using the homogeneous solution and the Local Density Approximation (LDA) \cite{LDH}.
Replacing the site index $j$ with a dimensionless variable $x$ the LDA assumes that each region at distance $x$
from the center of the trap can be well approximated  by a homogeneous system with
\be\label{lda}
\mu(x)=\mu(0)-\frac{m \omega_{e}^2}{2} x^2\, ,\ \ \ H(x)=H(0)\, ,
\ee
where $\mu(0)$ and $H(0)$ are the values of the chemical potential and magnetic field at the center
of the trap. For given values of $\mu(0), H(0)$ (or  equivalently $n(0)$ and $H$ with $n(0)$ the density
at the center of the trap) the density profiles can be computed using the QTM equations (\ref{qtm}) and
(\ref{lda}). Density profiles for the balanced system were previously investigated using quantum Monte Carlo
simulations \cite{RMBS}, density functional theory \cite{XPTCC,Campo} and DMRG \cite{FHM,HOF}.

We are going to investigate density profiles with $0<n(0)\le 2$ in the presence of a magnetic field. The phase diagram at zero temperature for both
negative and positive values of the chemical potential (corresponding to $n\in[0,2]$) is shown in Fig.~\ref{pddp}.
The properties of the phases at $\mu>0$ which are in one-to-one correspondence with  the five regimes analyzed in
Sect.~\ref{s21} can be determined using the symmetry relation (\ref{symm}) and Eqs.~(\ref{right}). For given values
of $U$, $m$ and $\omega_{e}$ we introduce a radius denoted by $R_{TF}$ and defined by
\be
2 \mu_0-\frac{m \omega_{e}^2}{2} R_{TF}^2=0\, ,
\ee
with $\mu_0=2(U/4)+2$ the value of the chemical potential for which the density in the
center of the trap is $n(0)=2$ at zero temperature and magnetic field. $R_{TF}$ is the distance from the center
of the trap at which the density profile is zero for a system in the ground state  with $n(0)=2$ and $H=0$.

In Fig.~\ref{densp} we present the density profiles for a system with
interaction strength $U=8$, $T=0.025$ and three values of the magnetic field
$H=\{0, 0.25, 0.5\}$ where $H=0.5>H_0(8)$. The temperature chosen is so low
that our results are almost indistinguishable from the $T=0$ case. The
structure of the density profiles is intricate, especially for $n(0)>1$, but
it can be easily understood by looking at the phase diagram from
Fig.~\ref{pddp}. Consider the case of $n(0)=2$ and $H=0.5$. Moving further
from the center of the trap (increasing $x$) is equivalent with the chemical
potential moving in the phase diagram from right to left along a horizontal
line (in this case the starting value of $\mu$ is the rightmost point on the
$H=0.5$ line of Fig.~\ref{pddp}). As $x$ increases from zero the system
crosses the following phases in order: $\mbox{I}_R$, $\mbox{II}_R$,
$\mbox{III}_R$, III, II and finally I. The total density profile can be
described as a band insulator at the trap center surrounded by metallic
regions in turn surrounded by Mott insulators \cite{LDH}. For $n(0)=2$ and
$H=0.25$ (the starting value of the chemical potential is the rightmost point
on the $H=0.25$ line) the system crosses the phases: $\mbox{I}_R$,
$\mbox{II}_R$, $\mbox{IV}_R$, $\mbox{V}_R$, V, IV, II and I. While the total
density profile is similar with the previous case (the metallic regions
present angular points at the $\mbox{II}_R\rightarrow\mbox{IV}_R$ and IV
$\rightarrow$ II transitions) the spin up and spin down density profiles are
very different as $H=0.5$ is larger than the critical field while $H=0.25$ is
below $H_0(8)$. Zero magnetic field is characterized by
$n_\uparrow(x)=n_\downarrow(x)=n(x)/2$ and it comprises the following phases
$\mbox{I}_R$, $\mbox{IV}_R$, $\mbox{V}_R$, V, IV, and I. The profiles with
$n(0)<2$ can be understood in the same way and noticing that in this case the
starting point of the chemical potential moves to the left which means that
the system can skip some of the phases enumerated in the $n(0)=2$ case. At
$n(0)=1.5$ the total density profiles can be characterized by a metallic
center surrounded by Mott insulator plateaus while for $n(0)=1$ we encounter
the situation of a Mott insulator at the center surrounded by metallic
wings. Finally, for $n(0)=0.8$ the system is completely metallic.  In all
cases the structure of the total density profile is similar for the same value
of the density at the center of the trap but the density profiles of the
components are heavily influenced by the presence of the magnetic field.

\section{Conclusions}\label{s10}

In this paper we have studied the influence of the magnetic field on the quantum critical behavior and
thermodynamic properties of the 1D repulsive Hubbard model. Even though all the QPTs investigated were
characterized by critical exponents $z=2$ and $\nu=1/2$ the universal thermodynamics in the vicinity
of the QCPs is not described by free fermions in all cases. The transitions from the vacuum belong to the
universality class of spin-degenerate impenetrable particle gas and the universal thermodynamics is given
by Takahashi's formula. The influence of the magnetic field on the thermodynamic properties is very important
at low temperature and small filling fractions. The magnetic susceptibility at zero magnetic field presents
an infinite slope at low temperatures for all values of the filling fractions.
The double occupancy exhibits two minima as a function of temperature in two situations:
a) at intermediate values of $U$ and b) for large values of the repulsion  and magnetic fields close to the
critical value. In all the other cases the double occupancy presents only one
minimum.
In the experimentally relevant
case of a trapped system the magnetic field does not influence strongly the overall density profiles but
changes dramatically the distribution of phases in the inhomogeneous system.
An interesting extension of our work is the case of the Hubbard model with impurities.
This will be addressed in a future publication.

\acknowledgments

O.I.P. acknowledges the financial support from the
LAPLAS 5 and LAPLAS 6 programs of the Romanian National
Authority for Scientific Research. A.K. is grateful to
DFG (Deutsche Forschungsgemeinschaft) for financial support
in the framework of the research unit FOR 2316. A.F. acknowledges
CNPq (Conselho Nacional de Desenvolvimento
Cientifico e Tecnologico) for financial support.

\appendix

\section{Numerical implementation of the first type of convolutions}\label{app1}

From the numerical point of view the main difficulty in the implementation of the
QTM integral equations (\ref{qtm}) is the treatment of the convolutions.
The first type of convolutions  is defined by
\be
K*f=\inti K(x-y) f(y)\, dy\, ,
\ee
where the kernels (defined in (\ref{kernels})) can be $K_{2,\pm\alpha-\alpha}(x)\, ,
K_{1,\pm \alpha}(x)$ or $\overline{K}_{1,\pm \alpha}(x)$ with  $ 0<|\alpha|<u\, $
 and $ f_\alpha(x)=f(x+i\alpha).$  The integrands are $\ln(1+\mathfrak{b}^\pm(x))$ or
$\ln(1+1/\mathfrak{b}^\pm(x)).$ In general the most efficient way of treating convolutions
is using the Fast Fourier Transform (FFT). However, in order to obtain accurate results
FFT requires that the functions are either periodic or they decrease rather fast at
infinity. In our case the kernels are relatively slowly decaying while the $\mathfrak{b}^\pm(x)$
functions have constant asymptotics at infinity, $\lim_{x\rightarrow\pm\infty}\ln
\mathfrak{b}^\pm(x)=-\beta H$. Denoting by $f(\infty)=\lim_{x\rightarrow\pm\infty}f(x)$ the
efficient way of treating this type of convolutions is by subtracting the asymptotic value
\begin{align}
\inti K(x-y) &f(y)\, dy =\inti K(x-y) [f(y)-f(\infty)]\, dy \nonumber\\
& + f(\infty)\inti K(x-y)\, dy\, .
\end{align}
In the first term on the right hand side now both the kernel and the integrand vanish
at infinity and can be calculated using FFT and an appropriate cutoff while the second
term can be analytically computed. We find
\begin{subequations}
\begin{align}
\inti K_{2,0}(x-y)\, dy&=1\, ,\\
\inti K_{2,2\alpha}(x-y)\, dy&=1\, ,\\
\inti K_{1,\alpha}(x-y)\, dy&=0\, ,\label{k1}\\
\inti K_{1,-\alpha}(x-y)\, dy&=1\, ,\label{k2}\\
\inti \overline{K}_{1,\alpha}(x-y)\, dy&=1\, ,\\
\inti \overline{K}_{1,-\alpha}(x-y)\, dy&=0\, .
\end{align}
\end{subequations}
Let us show how to compute these integrals. It is sufficient to consider
the case of
\[
I_1=\inti K_{1,\alpha}(x-y)\, dy
\]
all the other cases being amenable to a similar derivation. Making the change of variables
$x-y=z$ we obtain
\[
I_1=\inti\frac{u/\pi}{(z+ i\alpha)(z+i\alpha+ 2 i u)}\, dz \,  .
\]
First, we will consider the case of $\alpha>0$.
The integrand has two poles in the complex plane  situated at $z_1=-i \alpha$ and $z_2=-i\alpha-2iu$.
Closing the contour in the upper half plane by adding an infinite semicircle (which does not give
a contribution) the integral is analytic inside the closed contour and we obtain  $I_1=0$ proving (\ref{k1}).
In the case of $\alpha<0$ again closing the contour in the upper half plane we obtain  $I_1=2\pi i\,  \mbox{Res} (z=i|\alpha|)$
which gives $I_1=1$ proving (\ref{k2}).
All the other integrals can be computed in a similar fashion.

\section{Numerical implementation of the second type of convolutions}\label{app2}

The numerical implementation of the second type of convolutions defined by
\[
K\bullet f=\mbox{p.v.}\int_{-1}^{+1} K(x-y) f(y)\, dy\, ,
\]
is more complicated due to the presence of the principal value integral and also
because the integrands $\ln (1+\mathfrak{c}^\pm(x))$ and $\ln (1+\overline{\mathfrak{c}}^\pm(x))$
behave like $(1-x^2)^{1/2}$ in the vicinity of $\pm 1$. This weakly singular behavior suggest that the best way
to tackle these convolutions numerically is to employ Chebyshev quadratures.  First we will present some
minimal information on Chebyshev polynomials necessary to derive relevant quadrature formulae.

\subsection{Chebyshev polynomials}

We are going to use only Chebyshev polynomials of the first and second type defined by \cite{MH}
\begin{align}
T_n(x)&=\cos n\theta\, ,\ &x=\cos \theta\, ,\ \ \theta\in[0,\pi]\, ,\\
U_n(x)&=\frac{\sin[(n+1)\theta]}{\sin \theta}\, ,\ &x=\cos \theta\, ,\ \  \theta\in[0,\pi]\, .
\end{align}
The zeroes of $T_n(x)$ and $U_n(x)$ are given by
\begin{align*}
T_n(x):&\ x_k=\cos\left[\frac{(k-1/2)\pi}{n}\right]\, ,& \ \ k&=1,\ldots,n\, ,\\
U_n(x):&\ y_k=\cos\left[\frac{k\pi}{n+1}\right]\, ,& \ \ k&=1,\ldots,n\, \, .
\end{align*}
In the $z=\cos\theta$  variable the Chebyshev polynomials take the form
\begin{align}
T_n(x)&=\frac{1}{2}\left[(z+\sqrt{z^2-1})^n+(z-\sqrt{z^2-1})^n\right]\, ,\\
U_n(x)&=\frac{1}{2}\frac{(z+\sqrt{z^2-1})^n+(z-\sqrt{z^2-1})^n}{\sqrt{z^2-1}}\, .
\end{align}

\subsection{Chebyshev-Gauss quadrature}

In this section we remind the reader of the derivation of the  Chebyshev-Gauss quadrature
of the second kind which will constitute the basis for a similar result for  principal value integrals.
We want to obtain a quadrature for the integral
\be\label{a3}
I(f)=\int_{-1}^1 f(x)\, dx\, ,
\ee
where $f(x)\sim (1-x^2)^{1/2}$ at the endpoints of the interval but this requirement can be dropped.
We approximate $f(x)$ by a function of the form
\be\label{a4}
J_{n-1}[f(x)]=(1-x^2)^{1/2}\sum_{j=0}^{n-1} b_j U_j(x)\, ,
\ee
and we introduce
\be\label{a5}
I_{n-1}(f)=\int_{-1}^1 J_{n-1}[f(x)]\, dx\, ,
\ee
which constitutes an approximation of the integral (\ref{a3}) using an interpolating polynomial
of order $n-1$.
Using the second kind discrete orthogonality formula (8.33 of \cite{MH})
\[
d_{ij}=\sum_{k=1}^n(1-y_k^2)U_i(y_k)U_j(y_k)=\left\{\begin{array}{l} \frac{1}{2}(n+1)\, ,\ i=j\le n-1\, ,\\
                                                                     0\, , i\ne j
                                                    \end{array}                     \right.
\]
with $y_k=\cos(k\pi/n+1)\, , k=1,\ldots,n$ the zeroes of the Chebyshev polynomials of the second type
we obtain
\be\label{a6}
b_j=\frac{2}{n+1}\sum_{k=1}^n(1-y_k)^{1/2}f(y_k)U_j(y_k)\, .
\ee
Integrating (\ref{a4}) we find $I_{n-1}(f)=\sum_{j=0}^{n-1}b_ja_j$ with $a_j=\int_{-1}^1(1-x^2)^{1/2}U_j(x)\, dx.$
The coefficients $a_j$ can be calculated analytically with the result
\[
a_j=\int_0^\pi\sin [(j+1)\theta]\sin \theta\, d\theta=\left\{\begin{array}{l} \frac{\pi}{2}\, ,j=0\, ,\\
                                                                                0\, , j>0\, .
                                                               \end{array}                 \right.
\]
Collecting everything we find
\be\label{a7}
\int_{-1}^1 f(x)\, dx \sim I_{n-1}(f)=\sum_{k=1}^n\omega_k f(y_k)\, ,
\ee
with
\begin{subequations}\label{a8}
\begin{align}
\omega_k&=\frac{\pi}{n+1}\sin\left[\frac{k\pi}{n+1}\right]\, ,& \ \ k&=1,\ldots,n\, ,\\
y_k&=\cos\left[\frac{k\pi}{n+1}\right]\, ,& \ \ k&=1,\ldots,n\, .
\end{align}
\end{subequations}
Formulae (\ref{a7}) and (\ref{a8}) represent the Chebyshev-Gauss quadrature of the second kind
which are very efficient in the numerical integration of functions which behave like $(1-x^2)^{1/2}$
at the endpoints of $[-1,1]$.

\hspace{1cm}

\subsection{Chebyshev-Gauss quadrature for principal value integrals}

The same method can be used to derive a quadrature formula for principal value
integrals of the type
\be\label{a9}
I_\mathcal{C}(f)=\mbox{p.v.}\int_{-1}^1 \frac{f(x)}{y-x}\, dx\, ,\ \ |y|<1\, .
\ee
Using the same approximation (\ref{a4}) for the $f(x)$ function and repeating verbatim the
steps from the previous section we obtain $I_{\mathcal{C},n-1}(f)=\sum_{j=0}^{n-1}b_ja_j[y]$
with
\be
a_j[y]=\mbox{p.v.}\int_{-1}^1(1-x^2)^{1/2}\frac{U_j(x)}{y-x}\, dx=\pi T_{j+1}(y)\, ,
\ee
where the final result was derived using Theorem 9.1 of \cite{MH}. Together with (\ref{a6})
we find that $I_{\mathcal{C},n-1}(f)=\sum_{k=1}^n\omega_k(y)f(y_k)$ with
\[
\omega_k(y)=\sum_{j=0}^{n-1}\frac{2\pi T_{j+1}(y)}{n+1}\sin\left[\frac{k\pi}{n+1}\right] U_j\left[\cos\left(\frac{k\pi}{n+1}\right)\right]\, .
\]
Using $\sin\left[\frac{k\pi}{n+1}\right] U_j\left[\cos\left(\frac{k\pi}{n+1}\right)\right]=\sin\left[\frac{(j+1)k\pi}{n+1}\right] $ we
obtain the main result of this Appendix
\be
\mbox{p.v.}\int_{-1}^1 \frac{f(x)}{y-x}\, dx\, \sim I_{\mathcal{C},n-1}(f)=\sum_{k=1}^n\omega_k(y)f(y_k)\, ,
\ee
with
\begin{subequations}
\begin{align}
\omega_k(y)&=\sum_{j=1}^n\frac{2\pi}{n+1}T_j(y)\sin\left[\frac{(j+1)k\pi}{n+1}\right]\, , \\
y_k&=\cos\left[\frac{k\pi}{n+1}\right]\, ,\ \ \ \  k=1,\ldots,n\, .
\end{align}
\end{subequations}

\end{document}